\documentclass[fleqn,usenatbib]{mnras}

\usepackage[T1]{fontenc}

\DeclareRobustCommand{\VAN}[3]{#2}
\let\VANthebibliography\thebibliography
\def\thebibliography{\DeclareRobustCommand{\VAN}[3]{##3}\VANthebibliography}

\usepackage{newtxtext,newtxmath}
\usepackage{ulem}
\hypersetup{linkcolor=red,citecolor=blue,filecolor=cyan,urlcolor=magenta}
\usepackage{amsmath}
\usepackage{tensor}     
\usepackage{graphicx}   
\usepackage{bm}         
\usepackage{xcolor}     
\usepackage{color}      
\usepackage[section]{placeins} 
\usepackage[utf8]{inputenc} 
\usepackage[flushleft]{threeparttable}
\newcommand{\nc}{\newcommand*} 
\nc{\figurewidth}{3.2in}
\nc{\xbar}{\bar{x}}
\nc{\rhoeq}{\rho_{\mathrm{eq}}}
\nc{\zeq}{z_{\mathrm{eq}}}
\nc{\tla}{\tilde{\lambda}}
\nc{\dt}{\delta}
\nc{\Dt}{\Delta}
\nc{\vj}{\vec{j}}
\nc{\vl}{\vec{l}}
\nc{\hx}{\hat{x}}
\nc{\hy}{\hat{y}}
\nc{\bj}{\bm{j}}
\nc{\mJ}{\mathcal{J}}
\nc{\mP}{\mathcal{P}}
\nc{\Msun}{M_\odot}
\nc{\app}{\approx}
\nc{\av}[1]{\langle #1 \rangle}
\nc{\eq}[1]{Eq.~\eqref{#1}}
\nc{\al}{\alpha}
\nc{\Xstar}{X_{\ast}}
\nc{\seq}{\sigma_{\mathrm{eq}}}
\nc{\fpbh}{f_{\mathrm{pbh}}}
\nc{\vth}{\vec{\theta}}
\nc{\vla}{\vec{\lambda}}
\nc{\vd}{\vec{d}}
\nc{\Mmin}{M_{\mathrm{min}}}
\nc{\rmd}{\mathrm{d}}
\nc{\mmin}{{m_{\mathrm{min}}}}
\nc{\mmax}{{m_{\mathrm{max}}}}
\nc{\mR}{\mathcal{R}}
\nc{\tmR}{\tilde{\mathcal{R}}}
\nc{\s}{\sigma}
\nc{\ogw}{\Omega_{\mathrm{GW}}}
\nc{\addref}{[\textcolor{red}{add ref}] }
\nc{\Om}{\Omega}
\nc{\gpcyr}{\mathrm{Gpc}^{-3}\,\mathrm{yr}^{-1}}
\nc{\Eq}[1]{Eq.~\eqref{#1}}
\nc{\Fig}[1]{Fig.~\ref{#1}}
\nc{\Table}[1]{Table~\ref{#1}}
\nc{\lvc}{LIGO/Virgo} 
\nc{\Sec}[1]{Sec.~\ref{#1}}
\nc{\eg}{\textit{e.g.~}}
\nc{\SNR}{\mathrm{SNR}}

\def\({\left(}
\def\){\right)}
\def\[{\left[}
\def\]{\right]}

\def\e{\begin{equation}}
\def\q{\end{equation}}
\def\m{\begin{eqnarray}}
\def\n{\end{eqnarray}}

\title[Probing merger delay of SMBH binaries with GWs]{Probing the Delay Time of Supermassive Black Hole Binary Mergers With Gravitational Waves}

\author[Y. Fang $\&$ H. Yang]{Yun Fang,$^{1,2,3}$\thanks{E-mail: yunfang.phy@gmail.com}
Huan Yang,$^{1,4}$\thanks{E-mail: hyang@perimeterinstitute.ca}
\\
$^{1}$Perimeter Institute for Theoretical Physics, Waterloo, ON N2L2Y5, Canada\\
$^{2}$Kavli Institute for Astronomy and Astrophysics at Peking University, Beijing 100871, China\\
$^{3}$National Astronomical Observatories, Chinese Academy of Sciences, 20A Datun Road, Beijing, 100101, China\\
$^{4}$ University of Guelph, Guelph, Ontario N1G 2W1, Canada\\
}

\date{Accepted XXX. Received YYY; in original form ZZZ}
\pubyear{2022}
\begin{document}
\label{firstpage}
\pagerange{\pageref{firstpage}--\pageref{lastpage}}
\maketitle

\begin{abstract}
 Merging supermassive black hole binaries is expected as a consequence of galaxy mergers, yet the detailed evolution path and underlying merging mechanisms of these binaries are still subject to large theoretical uncertainties. In this work, we propose to combine the (future) gravitational wave measurements of supermassive black hole binary merger events with the galaxy merger rate distributions from galaxy surveys/cosmological simulations, to infer the delay time of binary mergers, as a function of binary mass.  The delay time encodes key information about binary evolution, which can be used to test the predictions of various evolution models.  With a Mock data set of supermassive black hole binary merger events, we discuss how to infer the distribution of delay time with hierarchical Bayesian inference and test evolution models with the Bayesian model selection method. The astrophysical model uncertainties are also considered in the hierarchical Bayesian inference and Bayesian model selection.
\end{abstract}


\section{Introduction}

Supermassive black holes (SMBHs) universally exist in galaxy centers. 
Observationally there are  fundamental relations between the mass of SMBHs and their host galaxy bulge luminosity ($M -$bulge) or stellar dispersion ($M-\sigma$)\citep{Dressler1988ApJ,Kormendy1993nag,Magorrian_1998,Ferrarese_2000,Gebhardt_2000,Kormendy:2013dxa}, which indicates the co-evolution between SMBHs and their host galaxies. 
It has been argued that the growth of SMBHs is related to the hierarchical mergers with other SMBHs, gas accretion and feedback, or the combination of both mechanisms in a way that they evolve towards the observed fundamental relations \citep{Volonteri_2003,DiMatteo:2005ttp,Hopkins_2006, Volonteri2009_Msigma}. The exact picture remains an open question, for which observations will play an important role in testing theoretical models. In particular, the mergers of SMBH binaries should happen as a consequence of galaxy mergers.  
In addition, there are substantial amount of evidences of galaxy mergers observed with electromagnetic observations \citep[see, e.g., ][for reviews]{Schweizer1996book, Lotz_2011,Conselice:2014joa}. The relevant key questions are how SMBHs form binaries and how the binaries evolve to merge, following the galaxy mergers. 

The dynamical evolution of two SMBHs after their host galaxy mergers depends on the mass ratio $\mu$ between the stellar mass of the progenitor galaxies, and the nuclear environment they reside in the descendant galaxy \citep{Begelman1980Natur, Dotti2012, Colpi2014, Merritt2005review, Vasiliev2016}. For major mergers ($\mu \in \[1/4,1\]$ with $\mu$ defined as the smaller mass divided by the larger mass), the dynamical friction is efficient to migrate the SMBHs to the center of the common nucleus of the descendant galaxies \citep{Callegari:2008py}. This process brings SMBH pair at a wide separation to a distance of 1-10 parsec within a timescale of about $\sim 10^8$yr \citep{Begelman1980Natur, Yu2002, Mayer:2007vk}, and then a hard binary forms where the mutual gravitational force between SMBHs becomes larger than the environmental gravitational force from stars or gas disc. 
For mergers with a small mass ratio, e.g., less than 1:10, the dynamical friction timescale could be larger than Hubble time and cause the secondary black hole to wander in the remnant galaxy \citep[see, e.g., ][]{Callegari_2011, Callegari:2008py}. 

After the hard binary is formed, the further decay of the SMBH binary orbit may be driven by the ejection of stars in the ``loss cone" region which has small angular momentum. In the nucleus of remnant galaxies which host spherical stellar distribution, the SMBH binaries tend to eject stars in the loss cone region rapidly and then form a depleted zone \citep{Milosavljevic_2002, Dullo_2014}.  
Since the re-population time scale of spherically distributed star clusters is usually longer than the Hubble time \citep{Yu2002}, the re-population efficiency is low so that further reduction of the binary separation is paused. 
At this stage, the GW radiation is however not strong enough to drive the binary to merge within Hubble time. It appears to be a problem to efficiently drive the SMBH binary across the parsec scale to the regime where GW radiation dominates. This is usually referred to as the ``final parsec problem" \citep{Milosavljevi__2001,Yu2002, Milosavljevic2003b}. 

On the other hand, there is circumstantial evidence that favors efficient SMBH binary coalescence. For example, the X-shaped radio sources (which are probably coalesced SMBH binaries with flipped jet directions \citep[][]{Dennett2002}) was observed to have a comparable rate to the expected rate of mergers of bright ellipticals, suggesting a quick coalescence following mergers \citep[][]{Merritt:2002hc}. In addition, if SMBH binaries fail to merge efficiently,  binaries with stalled orbit evolution should be present in many bright ellipticals, and subsequent galaxy mergers will bring in additional SMBHs to form multiple SMBH systems. The multiple SMBH system is observed to be rare, although the theoretical predictions suggest multiple SMBH systems may be common \citep{Hoffman_Loeb2007,Haehnelt2002}. 
The multi-body interaction may also slingshot eject SMBHs from the galactic centers, leading to much more scatter in the $M-\sigma$ and $M-$ bulge luminosity relations \citep[][]{Haehnelt2002}. 

To reconcile with observations, several mechanisms have been proposed to drive the orbital separation of SMBH binaries across the parsec scale. For example, a non-spherical shape of the galaxy, especially the triaxial-shaped gravitational potential gives rise to additional torques that change the angular momenta of stars, thus the loss cone remains full and the interaction with stars will continue to sink SMBH binary till the sub-parsec regime \citep[][]{Yu2002, Merritt_2004, Holley-Bockelmann:2006gbs, Gualandris2016}. 
Alternatively, in a gas-rich environment, the interaction of SMBH binary with the gaseous disk during the nuclear-disc-driven migration and later the binary-disc-driven migration will dissipate the orbital angular momentum and energy of the binary, thus provide another scenario that could efficiently drive SMBH binaries to GW-dominated regime  \citep[see, e. g., ][]{Armitage2002, Dotti_2006_nucleardisc, Haiman_2009}. We will discuss these two scenarios in Sec. \ref{sec_delay_models}, in terms of their observational implication with GW detection. Other proposed models include three body interactions of SMBHs as a result of multiple mergers \citep{Hoffman_Loeb_2007} and refilling of the loss cone via star-star encounters \citep[][]{Yu2002, Milosavljevic2003}. 

Different scenarios of the dynamical evolution of SMBH binary predict different delay times between galaxy mergers and the coalescence of SMBH binaries. The delay time is generally determined by the resident time the binary spends at their final-parsec stage, i.e., from several parsecs to sub-parsec separation before gravitational radiation takes over. To test the various dynamical models, observation of SMBH binaries at various evolution phases, especially at or below sub-parsec separation, is of key importance.

\begin{figure} 
\centering
\includegraphics[height=5.9cm]{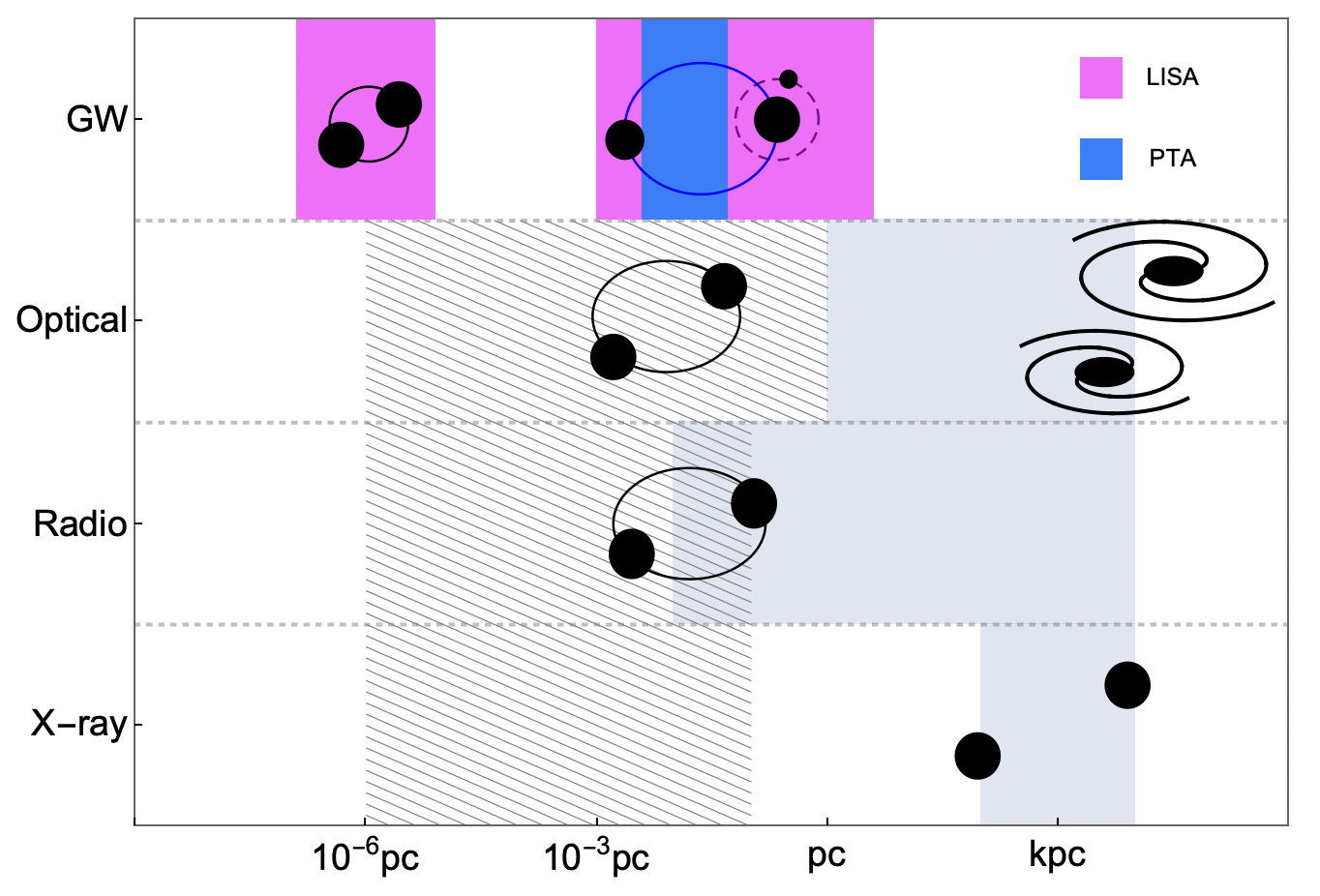}
\caption{The observations of SMBH pairs and binaries at their different evolution phases after galaxy mergers. The continuously filled regions in gray represent SMBH binaries/pairs that can be directly resolved by the EM telescopes \citep{Weisskopf_2000_Chandra,Perley_2011,Fu_2011,EventHorizonTelescope_2019I,Thatte_etal_2021,Breiding_2021}. The regions filled with lines represent the indirect detection of SMBH binaries with EM observations.
The pink-shaded region on the left represents the detection of SMBH binaries at their merger phases by LISA \citep{2017eLISA}, and the pink-shaded region on the right represents the measurement of SMBH binaries indirectly via EMRI waveform for which the dominant body of the EMRI is accompanied by a second SMBH \citep{Yunes:2010sm}. 
The blue-shaded region represents the GW detections with PTA over SMBH binaries at their inspiral phase \citep{Hobbs_etal_2010}, which is overlapped with LISA on the EMRI-SMBH binary measurements. 
}
\label{SMBHB_evolution_phase}
\end{figure}

Current searches for SMBH binaries are mainly performed with electromagnetic (EM) observations. The observation at X-ray and radio band have already revealed a few SMBH pairs at wide separations, from several parsec to kilo-parsec, by directly resolving two individual emission cores \citep[e.g., ][]{Komossa2003_NGC6240,Rodriguez:2006th, fu2011kiloparsec}. At smaller angular separation bellowing the limits of the resolution of EM telescopes,  SMBH binary candidates are indirectly found by diagnosing semi-periodic variations in the light curves \citep[e.g., ][]{Liu_2014,graham2015possible,Kharb_2017, Dey_2018, Britzen_2018,Bhatta_2019,Jiang_Yang_2022}, in the optical, X-ray, and radio waves. 
The detectability of SMBH binaries or SMBH pairs at their evolution phases with multi-band EM waves is schematically shown in Fig.~\ref{SMBHB_evolution_phase}.
The direct resolution of double cores at X-ray and radio band confirm the existence of SMBH pairs/binaries at wide separations. The optical spectroscopy observation helps to probe individual nuclei by searching for double-peaked emission lines, while the optical light curves alone are usually inconclusive to confirm the nature of emission sources, due to possible degeneracy with other origins \citep[see, e.g., ][]{fu2011kiloparsec, Wang_2010_Xray}. 
It is expected that the next generation of Event Horizon Telescope with better angular resolution may find sub-parsec SMBH binaries with both direct and indirect detections \citep{D_Orazio_2018,Fang:2021xab}.   

The gravitational waves (GWs) emitted by mergers of SMBH binaries are the most energetic astrophysical signals in the universe. 
The space-based gravitational wave detectors,  such as LISA (Laser Interferometer Space Antenna, \cite[][]{2017eLISA, eLISA:2013}) and Tianqin/Taiji \citep{luo2016tianqin,YueliangWu_taiji}, are designed to detect such events for binaries within the mass range $[10^3\Msun-10^8\Msun]$. It is predicted that LISA will be able to detect a few to several hundred GW events from SMBH binary mergers \citep[see, e.g., ][]{Klein:2015hvg}. The signals are so loud that LISA is able to detect some of the events up to a high redshift of $z\sim 20-30$\citep[][]{2017eLISA}. In addition to SMBH binary merger events, LISA will also observe a few to a few thousands of Extreme Mass Ratio Inspirals (EMRIs, which comprise a stellar-mass compact object orbiting around a SMBH) at low redshifts per year \citep{Babak:2017tow, Pan:2021ksp, Pan:2021oob}. It is suggested that \citep{Yunes:2010sm} if the host black hole of an EMRI system is a component of a SMBH binary, the massive companion at the separation range of $\[0.001\text{pc},10\text{pc}\]$
might be measured through the EMRI waveform which is modulated by the binary-induced center-of-mass motion. The fraction of EMRI systems in such SMBH binary environment may be enhanced due to an accelerated formation channel \citet{Mazzolari:2022cho,Naoz:2022rru}. 
We plot the detection of SMBH binaries at their merger phases by LISA in the left pink-shaded region of Fig.~\ref{SMBHB_evolution_phase}, and the indirect detection of SMBH binaries at the inspiral phase in the right pink-shaded region  of Fig.~\ref{SMBHB_evolution_phase}. 
At lower frequencies, Pulsar Timing Array (PTA, see, e.g., \cite{Hobbs_etal_2010}) is expected to detect  GWs from loud, individual SMBH binaries or the GW background (GWB) contributed by unresolved SMBH binaries within the mass range $\[10^8\Msun,10^9\Msun\]$ which are inspiraling at a separation about $0.01 \text{pc}$. 
Fig. \ref{SMBHB_evolution_phase} also plots the GW detections with PTA over SMBH binaries at their inspiral phase (blue-shaded region). 

Compared to EM observations, GW observations are often subjected to less observational systematics, and the selection bias is usually better understood. These features are vital for constructing an accurate description of SMBH binary populations. As a result, although in principle all observations mentioned in Fig.~\ref{SMBHB_evolution_phase} should be incorporated into a coherent framework to infer the evolution path and population of SMBH binaries, in practice it is technically challenging for such a task.  We shall only consider GW observations in this work. 
So far, the non-detection of GWB from SMBH binaries in PTA band has already constrained both the SMBH binary population \citep{Grahammnras1726, NANOGrav:2019tvo} and the individual candidates \citep{Jenet_2004}, although a sign of common process was found in the PTA data \citep{NANOGrav_12d5, Middleton:2020asl}. 
In the future, LISA will test the seeding channels and probe the evolution history of SMBHs/SMBH binaries in an unprecedented way \citep[e.g., ][]{Klein:2015hvg, Chen:2022sae, Toubiana:2021iuw}. 

We are interested in the delay time of SMBH binary coalescence after their host galaxy merge. By assuming the hierarchical merger model of SMBH formation and evolution following galaxy mergers, the merger rate of SMBH binary is determined by both the galaxy merger rate and the later time evolution histories. The galaxy merger rate could be observed from galaxy surveys \citep[e.g., ][]{Lotz_2011,Xu_2012,Mundy_2017, Casteels_etal_2014_GAMA, Driver_etal_2022GAMA, SDSS_merger_rate_2010, Davies_etal_DEVILS2018} or theoretically predicted from cosmological simulations \citep[e.g.,][]{Filip_2022_galaxy, OLeary_2021, Rodriguez-Gomez:2015aua}. 
In addition, the mass of SMBHs is observed to be statistically related to the properties of their host galaxies, as mentioned before. Based on these observations and predictions, one could infer the formation rate of SMBH pairs/binaries synchronously after galaxy merge. Finally, the delay time distribution is able to be inferred by associating the formation rate distribution of SMBH pair/binary with the merger rate distribution of SMBH binary, through the hierarchical Bayesian approach. 

LISA provides a promising opportunity for this task since it will provide the population information of SMBH binary merger events, such as the joint distribution of the redshift and mass etc., to a relatively high precision, thanks to its high single to noise ratio.  On the other hand, 
the delay time information is however unavailable from the stochastic gravitational wave background in PTA observation, due to the lack of detailed population information. 
If PTA also detect a set of SMBH binary merger events, they can also be used to infer the delay time. It is also worth to note that PTA generally observes SMBH binaries with larger masses  than those observed by LISA. 

In this work, we construct a mock data set of LISA GW events and show how to infer the delay time of SMBH binary coalescence after galaxy mergers based on the mock data. The delay time distribution is inferred via the method of hierarchical Bayesian inference, by combining the data of LISA GW events and galaxy merger rates (observed from galaxy surveys or predicted by cosmological simulations), together with the intrinsic relationship between SMBH mass and its host galaxy properties. We then compare different delay time models associated to various dynamical evolutionary scenarios, via the Bayesian model selection method.      

This paper is organized as follows. In Sec. \ref{sec_2}, we discuss the merger rate of SMBH binaries as a consequence of their host galaxy mergers. We construct the merger rate of SMBH binaries in Sec. \ref{sec_2.4}, based on the galaxy merger rate (Sec. \ref{sec_2.1}), the relation between SMBH binary mass and the host galaxy property (e.g., the stellar mass, in Sec. \ref{sec_2.3}), and the delay time distribution. Different delay time models associated to different dynamical scenarios are discussed in Sec. \ref{delay_stellar}--\ref{gas_rich_scenario}, and a phenomenological description of delay time distribution is given in Sec. \ref{delay_time_models}. 
In Sec. \ref{estimate_delay_distri}, we infer the delay time distribution from a mock LISA GW event set using the hierarchical Bayesian inference approach. Comparison between different delay time models associated to the corresponding dynamical scenarios are discussed in Sec. \ref{compare_delay_model}. The main results are discussed in Sec. \ref{conclusion}. In this article, we adopt the $\Lambda$ cold dark matter cosmology, with $\Omega_{M}=0.3$, $\Omega_{\Lambda}=0.7$, and $H_0=70\ {\text{km}}\ {\text{s}^{-1}\ \text{Mpc}}^{-1}$.


\section{Supermassive black hole binary mergers as a consequence of galaxy-galaxy mergers}
\label{sec_2}

According to the hierarchical merger model, SMBH binary merger events may happen as a consequence of galaxy-galaxy mergers. The merger rate of SMBH binaries is determined by the galaxy merger rate and the later-time binary evolution. The intrinsic relationship between SMBH mass and host galaxy properties also plays an important role in mapping the galaxy distributions to SMBH binary distributions. In this section, we construct the merger rate of SMBH binaries from the galaxy merger rate and the relationship between SMBH mass and host galaxy mass, together with a delay time distribution.    

\subsection{Galaxy merger rate}
\label{sec_2.1}

The galaxy-galaxy merger rate has been theoretically predicted based on semi-analytical models \citep[e.g.,][]{Filip_2022_galaxy}, semi-empirical models \citep[e.g.,][]{OLeary_2021}, and hydrodynamical simulations \citep[e.g.,][]{Rodriguez-Gomez:2015aua}. Observationally it can be obtained using the fraction of close galaxy pairs and the averaged merging timescale \citep[see, e.g., ][]{Lotz_2011, Xu_2012, Mundy_2017}. 
The merger timescale is still subject to large uncertainties and different approaches currently lead to non-converging predictions 
\citep[see, e.g., ][and the references therein]{Rodriguez-Gomez:2015aua,OLeary_2021,Filip_2022_galaxy}.
However, it is reasonable to expect that by the time the GW data from LISA is available, improved theoretical modeling together with galaxy surveys, e.g., the Galaxy And Mass Assembly Survey \citep[][]{Casteels_etal_2014_GAMA, Driver_etal_2022GAMA}, the Sloan
Digital Sky Survey \citep[][]{SDSS_merger_rate_2010}, and the Deep Extragalactic VIsible Legacy Survey \citep[][]{Davies_etal_DEVILS2018}, should provide a more consistent galaxy merger rate. 

In the following context, in order to illustrate how the delay time can be measured, we adopt the galaxy merger rate presented from the Illustris simulation by \cite{Rodriguez-Gomez:2015aua}, in which they have summarized a fitting expression of galaxy merger rate per galaxy, $\frac{{\rm d}N}{{\rm d}t} (z, M_*, \mu)$, as a function of redshift $z$, the stellar mass of the descendant galaxy $M_{*}$, and the stellar mass ratio $\mu$ of the progenitor galaxies (see Table \ref{table_galaxy_merger_rate} in the Appendix \ref{sec.app} for details). The galaxy merger rate density is given by the product of the galaxy merger rate per galaxy and the galaxy number density, i.e., the galaxy stellar mass function or Schechter function $\phi(z, M_{*})$, which can be written down as follows 
\m
{{\rm d}N_{\rm mergers} \over {\rm d}V_c \ {\rm d} \text{log}_{10}M_* \ {\rm d}t } = \frac{{\rm d}N}{{\rm d}t} (z,M_*,\mu) \times \phi(z,M_*) \,,
\n
where $\phi(z,M_*):={{\rm d}N_{\rm number} \over {\rm d}V_c \ {\rm d} \text{log}_{10}M_* }$, $N_{\rm mergers}$ is the number of galaxy mergers within the parameter bins, $V_c$ is the cosmological co-moving volume. 
The galaxy stellar mass functions in different redshift bins are taken from the recent observations given in \citep[][]{McLeod2021, Grazian2015AA, Stefanon_2021}. The details are shown in Table \ref{tab:Schechter_parameters}. 
In the future, it is straightforward to update this part with more accurate galaxy merger rate descriptions.
 
\subsection{Delay time models of supermassive black hole binary mergers}
\label{sec_delay_models}

The delay time of SMBH binary coalescence is closely related to the ``final parsec problem", as it is mainly determined by the dynamical evolution time  the SMBH binary spends to evolve from several parsec to sub-parsec separation. There are however different predictions of merger times considered in different dynamical scenarios, as suggested in various theoretical and numerical works  \citep[see, e. g.,][]{Yu2002, Milosavljevic2003, Mayer:2007vk, Haiman_2009}, or \citep[see][for reviews]{Merritt2005review, Colpi2014}. 
The observations of delay time through gravitational wave measurements will provide key insights into testing these models. In this work, we mainly focus on two scenarios for the matter of comparison, i.e., the stellar interaction and gas interaction scenarios, which are classified according to the dynamical evolution environment of the post-merger galaxies that the SMBH binaries reside in \citep[see, e. g., ][]{Begelman1980Natur, Yu2002, Milosavljevic2003, Mayer:2007vk, Haiman_2009, Antonini_2015}. 

\subsubsection{Delay time of SMBH binary in stellar environment}
\label{delay_stellar}
In a dry (negligible gas influence) major merger, the dynamical friction from the background stellar bulge of the post-merger galaxy is capable of bringing the two SMBHs to a bound orbit within a timescale $t_{\rm df}\lesssim 10 \text{Myrs} \ \text{to} \ 100\text{Myrs}$.
The remnant galaxy after a major merger is expected to be substantially nonspherical that may host triaxial potential. The centrophilic  orbits in triaxial galaxies can fill the loss cone of the phase space efficiently and harden the binary orbit towards the gravitational radiation stage. The hardening time scale is about a few $\text{Gyrs}$ that dominate the entire delay timescale in this scenario \citep[see, e. g., ][]{Yu2002,Khan_2011}. The detailed coalescence timescale after a bound binary SMBH is formed was estimated in \cite{Vasiliev_2015} using Monte Carlo
simulations, the results are as follows
\m \label{delaytime_stellar}
T_{\text{coal}} &\simeq& 1.7 \times 10^8 \text{yr} \times \left( r_{\text{infl}}\over 30 \text{pc} \right) ^{10+4\nu \over 5+\nu} \left( M\over 10^8 \Msun \right)^{- {5+3\nu \over 5+\nu}} \\ \nonumber 
&& \times {\xi}^{- {4\over 5+\nu}} \left( 4 q \over (1+q)^2 \right)^{3\nu -1\over 5+\nu} 20^{\nu} \\ \nonumber 
&& \times (1-e^2) \[ k+(1-k)(1-e^2)^4  \] \,,
\n
where $M$ and $q$ are the mass and mass ratio of SMBH binary, $k=0.4 + 0.1\ \text{log}_{10} (M/10^8 \Msun)$, and  the parameters $\nu$ and $\xi$ parameterize the  hardening rate with values estimated from Monte Carlo realizations (See Table 1 of \cite{Vasiliev_2015} for values of $\nu$ and $\xi$  in different types of triaxial galaxy). The influence radius of the binary SMBH $r_{\text{infl}}$ is approximated to be $r_{\text{infl}}=1.5 \ M/\sigma^2$ ( see \cite{Merritt_2006} for a more refined definition of $r_{\text{infl}}$). One may simplify Eq.~(\ref{delaytime_stellar}) by dropping the mass ratio dependent term $\left( 4q \over (1+q)^2 \right)^{3\nu -1\over 5+\nu}$, since it is almost unity for major mergers. Here, we assume no eccentricity dependence in delay time by taking circular orbit so that it reduce the complexity of the preliminary analysis. In fact, the eccentricity may play an important role in the delay time scale as shows in Eq.~(\ref{delaytime_stellar}) and also see \cite{Bonetti2019} for examples. 
By replacing $\sigma$ by $M$ according to the Faber–Jackson relation $M \propto \sigma^4 $\citep[][]{Faber_Jackson_1976, Lauer_etal_2007},  the coalescence time can be parametrized to be 
\m \label{coal_time_stallar_para}
T_{\text{coal}} &\simeq& \beta \left( M\over 10^6 \Msun\right)^{\alpha} \,,
\n
where $\beta$ and $\alpha$ depends on the undetermined values of $\nu$, $\xi$ and $e$. 

As the coalescence time $T_{\rm coal}$ generally dominates the lifetime of SMBH binary, we approximate the delay time in stellar environment by the coalescence time described in Eq.~(\ref{delaytime_stellar}) and (\ref{coal_time_stallar_para}):

\m
t_{\text{delay}, \text{stellar}} \sim T_{\text{coal}} \,. 
\n
Note that the power index of $M$, i.e. $\alpha$, only depends on $\nu$, and for all the listed values of $\nu$ in Table \ref{table_galaxy_merger_rate} of \cite{Vasiliev_2015},  $\alpha$ varies but always stays negative. For example, if we set $\xi=0.4$, $\nu=1/3$, and $e=0$, the coalescence time equals to 
\m \label{stellar_delay}
T_{\text{coal}} &\simeq& 1 \text{Gyr} \left( M\over 10^6 \Msun\right)^{-1/16} \,.
\n  

We can characterize the behaviors of delay time, with $\beta$ parametrizing the delay time scale and $\alpha$ parameterizing the dependence on the mass of SMBH binary.  These values may vary depending on the triaxial potential of the galaxy.  
\subsubsection{Delay time of SMBH binary in gas-rich environment} 
\label{gas_rich_scenario}
In gas-rich environments, the dynamical evolution is even more difficult to fully understand. We use a simplified delay time expression given in  \cite{Antonini_2015} which assumes the delay is simply controlled by the viscous timescale of the nuclear gas. The predicted delay time in this gas-rich nuclear environment is \citep{Antonini_2015, Granato_2004}
  \m \label{delay_time_gas}
  t_{\text{delay}, \text{gas}} \sim \mathcal{R}_c t_{\rm dyn} \,,
  \n
  where $\mathcal{R}_c$ is given by the critical Reynolds number in the range $\sim 10^2- 10^3$, and $t_{\rm dyn}$ is the dynamical time at the influence radius given by $t_{\rm dyn}= M/ \sigma^3$. Here we assume a benchmark value $\mathcal{R}_c = 10^3$ similar to \cite{Antonini_2015}. The delay time described by Eq.~(\ref{delay_time_gas}) is in the range $10\text{Myrs}$ to $ 100\text{Myrs}$, which is consistent with the studies of the coalescence time scale of SMBH binary in gas-rich environments \citep{Escala_2005,Dotti_2006_nucleardisc, Mayer:2007vk, Haiman_2009, Colpi2014}. 
  
By replacing $\sigma$ again with $M$ using the Faber–Jackson relation $M \propto \sigma^4 $, we  rewrite Eq.~(\ref{delay_time_gas}) to 
\m \label{gas_delay}
 t_{\text{delay}, \text{gas}} \sim 40 \ \text{Myr} \times \left({M\over 10^6 \Msun} \right )^{1/4} \,.
\n

It is also worth noting that other models of gas-rich scenarios predict different delay times. For example, \cite{Goicovic:2016dul} considered the dynamical evolution of SMBH binaries interacting with infalling gas clumps by exchanging angular momentum through gas capture and accretion. \cite{Goicovic:2016dul} predicted a dynamical timescale of $0.1\text{Gyr}$ to $1\text{Gyr}$ for SMBH binaries to evolve into the GW emission regime. 

Similar to the stellar scenario, we also phenomenologically parametrize the delay time in gas-rich scenario as 
\m
 t_{\text{delay}, \text{gas}} \sim \beta \left({M \over 10^6} \right)^{\alpha} \,.
\n

In this way, the delay time $t_{\text{delay}, \text{gas}}$ and $t_{\text{delay}, \text{stellar}}$ predicted by different dynamical scenarios are both parameterized with $\alpha$ and $\beta$. 

\subsubsection{Phenomenological delay time distribution models}
\label{delay_time_models}

The delay time in the stellar dynamical scenario and gas-rich scenario described in Eq.~(\ref{stellar_delay}) and Eq.~(\ref{gas_delay}) share the same form while differed by the underlying values of $\alpha$ and $\beta$. 
In reality, the astrophysical environments SMBH binaries reside in are complicated so that the asymmetry of stellar potential, the gas friction, and/or other properties of the host galaxies may play different roles in the dynamical evolution of SMBH binaries \citep[e.g., ][]{Mayer:2007vk, Callegari_2011, Vasiliev_2015, Chen_2020}. To describe the statistical distribution of delay times in various galaxies, we use two phenomenological models for the delay time distribution: the Gaussian delay and the Power-law delay.   

{\it Gaussian delay distribution.} In the ``Gaussian" delay distribution model, we parameterize the delay time distribution in the following way:  
\m \label{gaussian_delay}
P_{\rm delay} (\tau |M, {\bf \Lambda}) = N( u, \sigma ) {2 \over 1+\text{erf}({ u \over \sqrt{2} \sigma }) }\,,
\n
where $N( u, \sigma)$ denotes a Gaussian normal distribution with a mean value $ u=\beta ({M\over 10^6 \Msun})^{\alpha}$ and a standard deviation $\sigma$, and ${\bf \Lambda}$ here denotes $(\alpha,\beta,\sigma)$ which parameterize the delay distribution model in this case. The delay time $\tau$ takes the range $\[0, \infty\]$, so that a normalization constant ${2 \over 1+\text{erf}({u \over \sqrt{2} \sigma }) }$ in Eq.~(\ref{gaussian_delay}) is required to keep the total probability unity.

 {\it Power-law delay distribution.} In the Power-law delay model, we parameterize the delay time distribution to be 
\m \label{powerlaw_delay}
P_{\rm delay} (\tau|M, {\bf \Lambda})  \propto {1\over (\tau/\text{Gyr})^{\gamma \ \text{log}_{10} M + \kappa} }\,,
\n
with $\tau$ takes value between $\[\tau_{\text{min}}, \tau_{\text{max}}\]$, where $\tau_{\text{min}}$ and $\tau_{\text{max}}$ are the lower and upper limit of delay time respectively, and here ${\bf \Lambda}=(\gamma, \kappa)$. 
We set $\tau_{\text{min}}=0.001\text{Gyr}$ which is consistent with the dynamical friction timescale. The maximal delay time $\tau_{\text{max}}$ could in principle be larger than Hubble time, corresponding to the case that SMBH binary never merger. 
In that case, we can set $\tau_{\text{max}}$ to an arbitrarily large number since the contribution from the tail in the power law distribution is small and negligible. Therefore we assume $\tau_{\text{max}}=30\text{Gyr}$. 
  
  
  
  

\subsection{Relationship between Supermassive black hole mass and host galaxy property}
\label{sec_2.3}

The measurement of delay time distribution requires the comparison between the formation rate distribution of SMBH pairs/binaries (synchronously after galaxy mergers) and the merger rate distribution of SMBH binaries (through observation with GWs). The SMBH pair/binary formation rate relies on the galaxy merger rate and the fundamental relationship between the SMBH mass and the host galaxy properties, such as the $M$--bulge mass or the $M$--stellar velocity dispersion relations. In this work, we shall assume the relationship between the binary mass $M$ and the stellar mass $M_{*}$ of the descendant galaxy, i.e., the $M$--$M_*$ relationship. Observationally, the $M$--$M_*$ relationship is under certain divergence at different mass bins and subjected to uncertainties at high redshift bins \citep[e.g., ][]{Kormendy:2013dxa, Reines_2015, Ding_etal_2020}. For a simple illustration purpose, we shall take the $M$--$M_*$ relation from the observations of bulge dominated galaxies (where $M_{*} \sim M_{\text{bulge}}$), which gives \citep{Kormendy:2013dxa}:   
\m \label{M_Mstar_relation}
\text{log}_{10}M= a + b\ \text{log}_{10}\left ({M_{*} \over 10^{11}\Msun} \right ) \,,
\n
where $a=8.69 \pm 0.05$, $b=1.17 \pm 0.08$, and the intrinsic scatter is $\epsilon=0.28 \ \text{dex}$. 

\subsection{Supermassive black hole binary merger rate} 
\label{sec_2.4}
With the galaxy merger rate, the delay time model, and the relationship between the SMBH binary mass and stellar mass of descendant galaxy presented in Sec. \ref{sec_2.1}-- \ref{sec_2.3}, we now construct the merger rate distribution of SMBH binaries as follows 
  \m \label{BH_merger_rate1}
 \mathcal{R}(M, t_{\text {L}} | {\bf \Lambda})=  \int  R_g [t_{\text {L}} + \tau, X(M)] {dX\over dM} P_{\text{delay}} (\tau|M, {\bf \Lambda}) d\tau, 
  \n
  where $t_{\text{L}}$ is the lookback time of  SMBH binary merger, $R_g$ is the galaxy merger rate, and $X$ either denotes the stellar mass $M_{*}$, the bulge mass $M_{\text{bulge}}$, or the bulge luminosity $L_{\text{bulge}}$ of the host galaxy, which is a function of $M$ according the observed fundamental relationships. If the uncertainty of such relationship (mainly from the intrinsic scatter) is taken into account, Eq.~(\ref{BH_merger_rate1}) can be rewritten as
  \begin{align} \label{BH_merger_rate2}
\mathcal{R}(M, t_{\text {L}} | {\bf \Lambda})\!\! =\!\!\!  \int  \!\!\!\, R_g (t_{\text {L}} \!\!+\!\! \tau, X)  P_{\text{delay}} (\tau|M, {\bf \Lambda}) P (M| X) d X d\tau, 
  \end{align}
  where the conditional probability $P(M|X)$ can be obtained through the joint distribution between $M$ and $X$ through $P(M,X)/P(X)$.
  
  We take the galaxy merger rate $R_g (t_{\text{L}}, M_{*})$ in Eq.~(\ref{BH_merger_rate1}) (and (\ref{BH_merger_rate2})) as a marginalized distribution over mass ratio within the range $\[1/4,1\]$:  
\\
\m \label{galaxy_merger_rate_t_Mg}
R_g (t_{\text{L}}, M_{*})&:=&{{\rm d}N_{\rm mergers} \over {\rm d}z \ {\rm d} \text{log}_{10}M_* \ {\rm d}t} \times { {\rm d}z\over {\rm d} t_{\text{L}}} \Big|_{z=z(t_{\text{L}})} \, \\ \nonumber
&=& \frac{{\rm d}N}{{\rm d}t} (z, M_*) \times \phi(z,M_{*}) \times {{\rm d}V_c \over {\rm d} t_{\text{L}} } \Big|_{z=z(t_{\text{L}})} \,,
\n
where 
\m
\frac{{\rm d}N}{{\rm d}t} (z, M_*)=\int^{1}_{0.25} \frac{{\rm d}N}{{\rm d}\mu \, {\rm d}t} (z, M_*, \mu)\ {\rm d} \mu \,,
\n
as here we are considering major mergers.  One can infer a similar range of mass ratio for SMBH binaries from the $M-M_{*}$ relation. We neglected the explicit dependence on the mass ratio of SMBH binary in the distribution  based on several reasons. Firstly, for the delay time distribution determined by the interaction of SMBH binary with stars, such as the case described by Eq.~(\ref{delaytime_stellar}), the delay time only weakly depends on the mass ratio of the binary. Secondly, for the delay time predicted in gas-rich environment via the nuclear-disc-driven migration  \citep{Dotti_2006_nucleardisc} or/and the binary-disc-driven migration \citep{Haiman_2009}, as has been discussed in Sec. \ref{gas_rich_scenario} with a simplified description, it also shows weak dependence on mass ratio in the case of major merger. Thirdly, physically major mergers are more successful to form compact/hard SMBH binaries \citep[e.g.,][]{Callegari_2011, Callegari:2008py, Colpi2014}. Finally, including the data of mass ratio is more computationally expensive. Since we use a interpolation function to replace the merger rate distribution (given in the form of convolution) in Eq.~(\ref{SMBHB_merger_rate_integ})
 to estimate the delay time distribution via the method of hierarchical Bayesian inference, it requires computationally more expensive process to obtain the high dimensional interpolation function if we include the mass ratio as an extra dimension. 


In this way, we constructed the SMBH binary merger rate $\mathcal{R}( M, z | {\bf \Lambda})$ by linking to the galaxy merger rate through the $M$-$M_{*}$ relationship and a delay time distribution, as   
 \m \label{SMBHB_merger_rate_integ}
 \mathcal{R}( M, z | {\bf \Lambda}) &=& \mathcal{R}( M, t_{\text{L}} | {\bf \Lambda})  { {\rm d} t_{\text{L}} \over {\rm d}z }|_{t_{\text{L}}=t_{\text{L}}(z)}  \\ \nonumber
 &=&  \int_{\tau_{\text{min}}}^{\tau_{\text{max}}} \! R_g (t_{\text{L}} \! + \! \tau, M_*) P_{\text{delay}} (\tau|M, {\bf \Lambda}) d\tau \\ \nonumber
 &&\times \ { {\rm d} \text{log}_{10}M_*  \over {\rm d} \text{log}_{10}M } \ {{\rm d} t_{\text{L}} \over {\rm d}z }\Big|_{t_{\text{L}}=t_{\text{L}}(z)}  \\ \nonumber
 %
 %
 \n
  where we have taken $X$ as $M_*$. 

Before we move onto detailed calculations, we notice that the stellar mass functions given in Table \ref{tab:Schechter_parameters} are discontinuous piecewise functions fitted from observations at different redshift bins, which produce discontinuities in SMBH binary merger rates constructed with Eq.~(\ref{SMBHB_merger_rate_integ}), and also lead to auxiliary oscillations when  the integration in Eq.~(\ref{SMBHB_merger_rate_integ}) is performed. To improve on these artificial issues, we smooth out the galaxy stellar mass function given in Table \ref{tab:Schechter_parameters} by averaging the stellar mass function in the transition regions between two  neighbor bins.  

\begin{figure} 
\centering
\includegraphics[height=6.5cm]{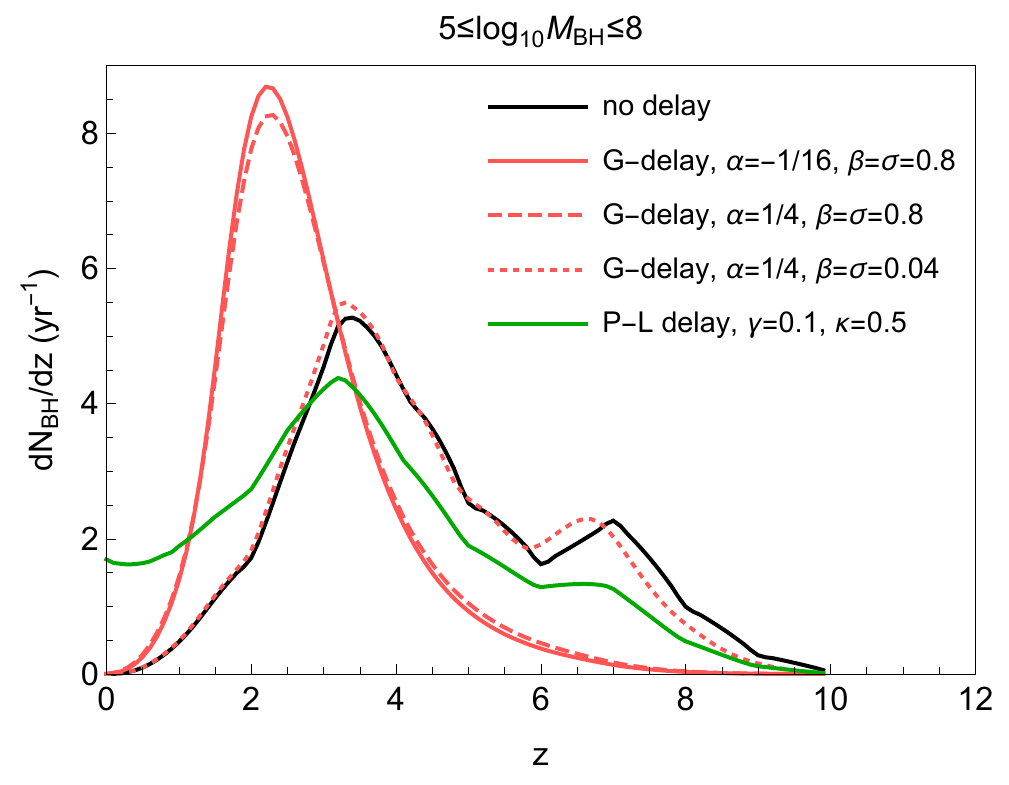}
\caption{The marginalized SMBH binary merger rate over mass as a function of redshift given by different delay time models. The SMBH binary merger rate is derived from Eq.~(\ref{SMBHB_merger_rate_integ}), considering the best fit value of the galaxy merger rate (Table~\ref{table_galaxy_merger_rate} and Table~\ref{tab:Schechter_parameters}), and the best fit of the relationship between the binary mass and host galaxy stellar mass (Eq.~(\ref{M_Mstar_relation})). The delay time models are phenomenologically parameterized with a Gaussian function or a Power-law function given in Eq.~(\ref{gaussian_delay}) and Eq.~(\ref{powerlaw_delay}). The legend for different line styles denotes the delay parameters for different delay models.}
\label{RBH_z_models}
\end{figure}

The SMBH merger rate for several sample delay models is shown in Fig. \ref{RBH_z_models}. 
In particular, the SMBH merger rate is marginalized over mass to  be a function of redshift, in which we take the best fit of galaxy merger rate given in Table \ref{table_galaxy_merger_rate} and \ref{tab:Schechter_parameters}, the $M$--$M_{*}$ relation given in Eq.~(\ref{M_Mstar_relation}), and assume four different delay models. The first three delay models are ``Gaussian" delay distribution (described in Eq.~(\ref{gaussian_delay})) parameterized by three groups of parameters $(\alpha,\beta,\sigma)$. 
In the case of the solid red line, we take $\alpha=-1/16$ and $\beta=\sigma=0.8 \text{Gyr}$ to represent the stellar environmental delay scenario discussed in Sec. \ref{delay_stellar}; In the dotted red line case, we assume $\alpha=1/4$ and $\beta=\sigma=0.04 \text{Gyr}$ to represent the gas-rich delay scenario discussed in Eq.~(\ref{gas_delay}); In the dashed red line, we take $\alpha=1/4$ and $\beta=\sigma=0.8\text{Gyr}$ to represent either stellar scenario with different mass dependence or gas-rich scenario with longer delay timescale as predicted in \cite{Goicovic:2016dul}. 
As a comparison, we also consider a delay model with Power-law delay distribution (described in Eq.~(\ref{powerlaw_delay})), as shown in the green line. 

 From Fig. \ref{RBH_z_models} we see that the ``Gaussian" delay model with delay timescale of  $\mathcal{O}(\text{Gyrs})$ (solid and dashed red lines) induces an obvious change to SMBH merger rate compared to the case without delay (black line), by shifting the position of redshift where the peak locates to a smaller value, and simultaneously raising the amplitude of the peak. The case with delay time of order $\mathcal{O}(10 \text{Myrs})$ (dotted red line) lead to an almost indistinguishable merger rate distribution as the one without delay. The Power-law delay model, such as the one shown in the green line, tends to broaden the distribution of merger rate towards smaller redshift while keeping the peak position less affected. The number of merger events predicted by these models is about 18 ${{\text{year}}^{-1}}$ -- 20 ${{\text{year}}^{-1}}$, which is consistent with the previous results \citep[e.g.,][]{Klein:2015hvg, Bhagwat:2021kwv}.   
The variation of predicted binary merger rates in different models is small because the delay time scales from these delay models are approximately between $\mathcal{O}(1)$Gyr to $10$Myrs, so that most of the systems merge  within Hubble time. The merger number per year is therefore close to the result without delay, as predicted under the same underlying galaxy merger rate and $M$--$M_*$ relation (which gives 19.5 $\text{year}^{-1}$).  

\section{Estimating delay time distribution using Hierarchical Bayesian inference}
\label{estimate_delay_distri}

The discussion in Sec.~\ref{sec_2} shows how the SMBH binary merger distribution is influenced by the underlying delay time distribution. In particular, under the same assumption of the galaxy merger rate and the $M$--$M_*$ relationship, the SMBH binary merger distribution displays rather distinctive features as shown in Fig. \ref{RBH_z_models}. Thus, the delay time information should be encoded in the merger distribution of SMBH binaries, which can be inferred from the difference in the distribution between the SMBH binary merger rate and the formation rate of SMBH pairs/binaries. In this section, we discuss how to use the set of GW events to infer the underlying delay time distribution. For this purpose, we shall apply the framework of Hierarchical Bayesian Inference. 

\subsection{Population analysis framework}
\label{sec_frame}

The hierarchical Bayesian approach is commonly used to infer the underlying distribution with a set of events statistically following such distribution.
The underlying distribution is usually parameterized with a mathematical expression with a few hyperparameters, which are constrained by the observed data using Bayesian Inference. This method has been previously applied to infer the population properties of binary black holes with LIGO/Virgo events from GWTC \citep[][]{Abbott_2021_GWTC2,LIGO_2021_GWTC3}.

In this section we infer the distribution of delay time of SMBH mergers with mock data sets of LISA GW events via hierarchical Bayesian approach, following the similar way that the binary black hole population property is inferred with LIGO/Virgo events. For the SMBH binaries considered here, the population distribution of LISA events is determined by the SMBH pair/binary formation rate combined with a delay time model, where the SMBH pair/binary formation rate is inferred from the galaxy merger rate (from observations or theoretical predictions) together with the relationship between the SMBH binary mass and host galaxy properties (from observation). The total number of events is modeled as an inhomogeneous Poisson process, and the joint likelihood function is given by 
\m \label{def_hyper_likelihood}
\mathcal{L}(\{{\bf d}\} | {\bm \Lambda} ) &\propto& {N({\bf \Lambda})}^{N_{\text{det}}} e^{- N_{\text{exp}} ({\bf \Lambda}) }  \times  \\ \nonumber
&& \prod^{N_{\text{det}}}_{i=1} \int \mathcal{L}(d_i |{\bm \theta}) P ({\bm \theta}| {\bm \Lambda}) d {\bm \theta}
\n 
 where $N_{\rm det}$ is the number of detected events, $N_{\text{exp}}$ is the expected number of detections within the observation duration for the population model ${\bm \Lambda}$ ( $\bm \Lambda$ represents the hyperparameters of the given model),  ${\bm \theta}=(m_1, m_2, z, ...)$ represents source parameters, $N({\bf \Lambda})$ is the predicted number of merger events under the model ${\bf \Lambda}$, $N_{\text{exp}}=\xi({\bf \Lambda})  N $ represents the expected detection number assuming a detection fraction $\xi({\bf \Lambda})$, $\mathcal{L}(d_i |{\bm \theta})$ is the likelihood function for each individual event, and $P ({\bm \theta}| {\bm \Lambda})$ denotes the population distribution corresponding to model $\bm \Lambda$. 
%


The integral in Eq.~(\ref{def_hyper_likelihood}) can be evaluated by generating Monte-Carlo (MC) posterior samples ${\bm \theta}_i$ following the likelihood function  $\mathcal{L}(d_i |{\bm \theta})$ in the parameter estimation process for each event. The expression can be rewritten as
\m \label{redef_likelihood}
\mathcal{L}(\{{\bf d}\}|{\bm \Lambda}) \propto  
{N({\bf \Lambda})}^{N_{\text{det}}} e^{- N_{\text{exp}} ({\bf \Lambda}) } 
\prod^{N_{\text{det}}}_{i=1} \langle {P ({\bm \theta}| {\bm \Lambda}) \over P_{\varnothing} ({\bm \theta})} \rangle
\n
where $ \langle ... \rangle$ is the average over the Monte-Carlo samples ${\bm \theta}_i$ of the individual events, and here $P_{\varnothing} ({\bm \theta})$ is the default prior taken in the parameter estimation. The detection fraction $\xi({\bm \Lambda})$ here is taken to be one as we simply assume all the SMBH binary merger events within the $10^5-10^8 M_\odot$ mass range (the range of mass considered in the mock data set) are detectable by LISA. This assumption is  based on the fact that LISA can detect SMBH binary mergers (with comparable mass) over a wide mass bin and across a large redshift \citep{2017eLISA}. 

The posterior of the hyperparameters, $P ({\bm \Lambda} | \{ {\bf d} \})$, given the the population model ${\bm \Lambda}$ and data $\{{\bf d}\}$ is  
\m
P ({\bm \Lambda} | \{ {\bf d} \}) \propto \mathcal{L}(\{{\bf d}\}|{\bm \Lambda}) P ({\bm \Lambda})\,.
\n

We consider the  population properties of binary mass $M$ and redshift $z$, i.e., ${\bm \theta}=(M,z)$. 
The population distribution $P({\bm \theta}| {\bm \Lambda})$ is the normalized  merger distribution of $ \mathcal{R}( M, z| {\bf \Lambda})$ (Eq.~(\ref{SMBHB_merger_rate_integ})), defined as 
\m \label{population_prob}
P({\bm \theta}| {\bm \Lambda}) =  {1\over \mathcal{N}({\bf \Lambda})}\mathcal{R}( M, z | {\bf \Lambda}) \,,
\n
where 
\m
\mathcal{N}({\bf \Lambda})=\int {\rm d} \text{log}_{10}M \ \text{d}z \ \mathcal{R}( M, z | {\bf \Lambda})
\n
is the merger events per year of SMBH binaries. The predicted merger number $N({\bf \Lambda})$ is $\mathcal{N}({\bf \Lambda})$ times the  duration of observation $T_{\text{det}}$, $N({\bf \Lambda})= \mathcal{N}({\bf \Lambda}) \ T_{\text{det}}$. 

The hierarchical Bayesian inference analysis requires a definite distribution model of SMBH binary merger rate, which in turn requires a definite result of galaxy merger rate and the $M$--$M_*$ relation. In reality, the latter two components are still subject to large uncertainties, either from observations or from theoretical predictions. 
To account for the influence of the uncertainties, we  further extend the Likelihood in Eq.~(\ref{def_hyper_likelihood}) to a statistically averaged expression in which the uncertain fitting parameters satisfy their corresponding distributions. We denote ${\bm \lambda_1}$ as the fitting parameters in the galaxy merger rate, such as the parameters listed in Table \ref{table_galaxy_merger_rate} and \ref{tab:Schechter_parameters} for our assumption of galaxy merger model. Similarly, we use ${\bm \lambda_2}$ to denote the fitting parameters in the $M$--$M_*$ relationship, such as the one taken in Eq.~(\ref{M_Mstar_relation}). Then the averaged Likelihood takes the form 
\m \label{redef_likelihood_errors}
\!\!\! \! \bar{\mathcal{L}}(\{{\bf d}\}|{\bm \Lambda}) &=& \! \int \!  \mathcal{L}(\{{\bf d}\}|{\bm \Lambda}, {\bm \lambda_1}, {\bm \lambda_2}) P({\bm \lambda_1}) P( {\bm \lambda_2}) d {\bm \lambda_1} d {\bm \lambda_2} \nonumber \\
&=& \sum_{{\bm \lambda_1},{\bm \lambda_2}} \mathcal{L}(\{{\bf d}\}|{\bm \Lambda}, {\bm \lambda_1}, {\bm \lambda_2}) \,,
\n
where $\mathcal{L}(\{{\bf d}\}|{\bm \Lambda}, {\bm \lambda_1}, {\bm \lambda_2})$ is the likelihood inherited from Eq.~(\ref{redef_likelihood}) associated with the galaxy merger rate $R_{\text{g}}(t_{\text{L}},M_{*} | {\bm \lambda_1})$ and the relationship $M_{*}(M |{\bm \lambda_2})$, throuth Eq.~(\ref{population_prob}) and (\ref{SMBHB_merger_rate_integ}). The summation in the second equal is the average over parameters ${\bm \lambda_1}$ and ${\bm \lambda_2}$ which satisfy the corresponding probability distribution functions (PDFs) $P({\bm \lambda_1})$ and $P({\bm \lambda_2})$.

The best fit and error bars of the fitting parameters are given in Table \ref{table_galaxy_merger_rate}, \ref{tab:Schechter_parameters}, and Eq.~(\ref{M_Mstar_relation}). We use  Eq.~(\ref{M_Mstar_relation}) to describe the SMBH-galaxy relationship, where the main uncertainty comes uniquely from the intrinsic scatter. However, this uncertainty has been absorbed in the convolution of Eq.~(\ref{BH_merger_rate2}), which replaces the galaxy's stellar mass with the mass of SMBH. The error bars for the rest of the parameters in ${\bm \lambda_2}$, i.e., $a$ and $b$ are much smaller compared to the error bar for the scatter and thus we could simply replace $a$ and $b$ with the best-fit values. The remaining consideration of uncertainties is for  ${\bm \lambda_1}$. We construct the PDFs for the parameters in ${\bm \lambda_1}$ either with a Gaussian/Gamma distribution or with a composition of two different Gaussian that are consistent with the corresponding best fit and error bar values. The details of how we generate these PDFs are given in the Mathematica notebook in \cite{fang_SMBHB_delay}. 

\subsection{Estimating the delay time distribution}
\label{estimate_delay_posteriors}

\begin{figure} 
\centering
\includegraphics[height=6cm]{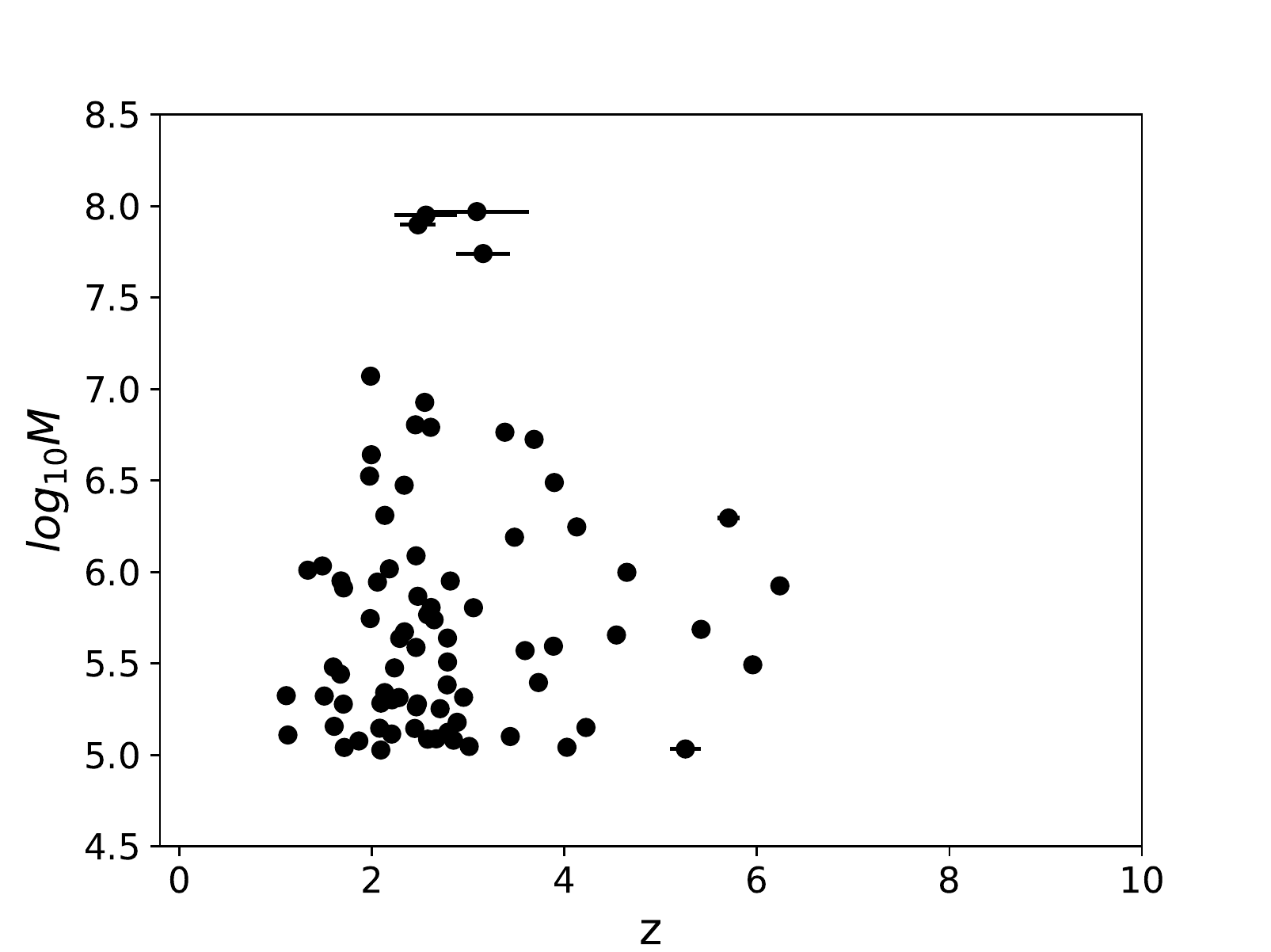}
\caption{The samples of mock GW data assuming four years observation period of LISA, obtained from our ``Gaussian" delay model with $\alpha=-1/16$ and $\beta=\sigma=0.8\text{Gyr}$. The black points represent the total 78 GW events denoted by source parameters $M$ and $z$. The error bars of $M$ and $z$ in each GW event are estimated from the Fisher information matrix according to the LISA sensitivity curve. }
\label{data_samples_Gaussian_delay}
\end{figure}

\begin{figure} 
\centering
\includegraphics[height=6cm]{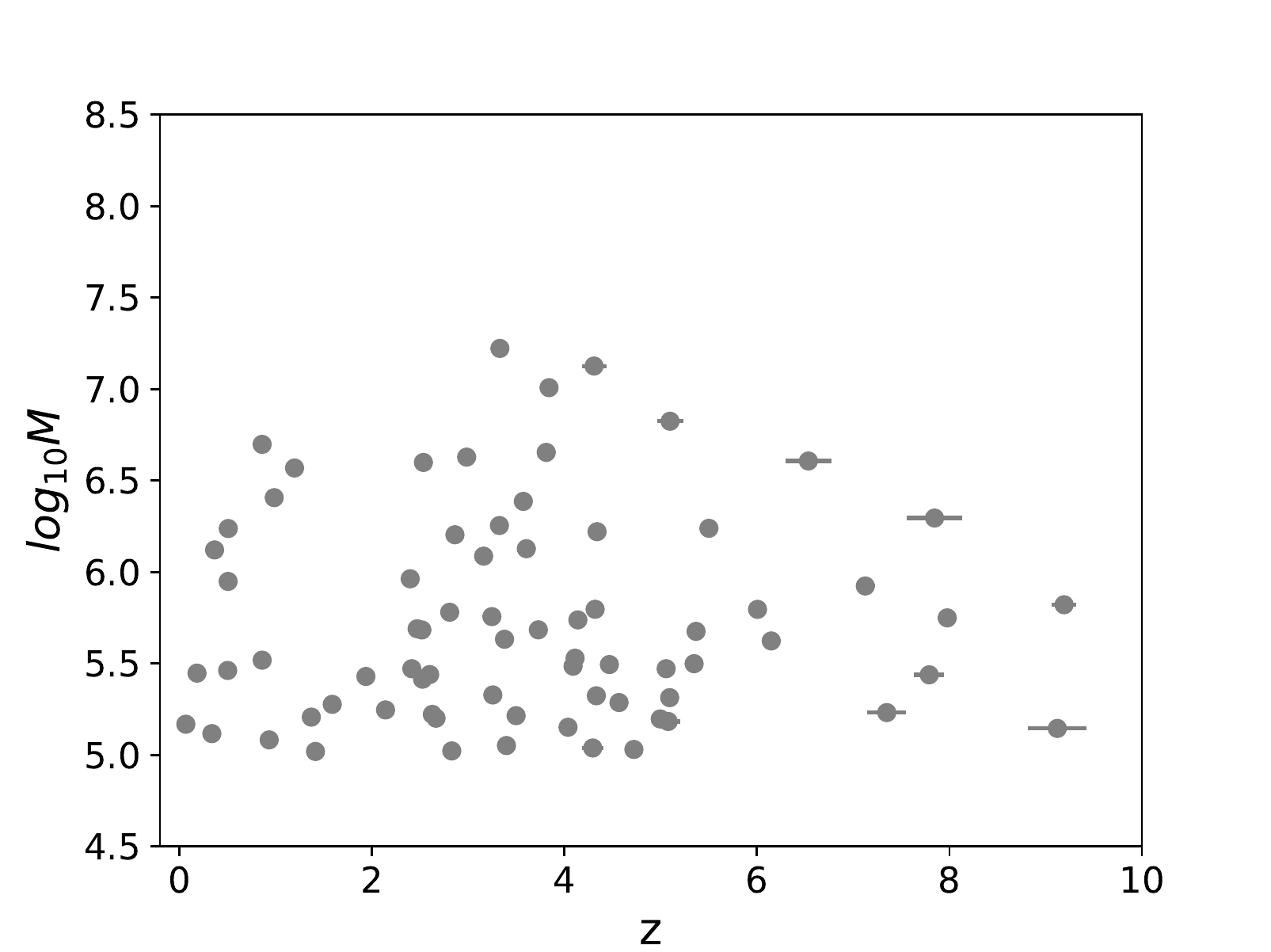}
\caption{Similar to  \ref{data_samples_Gaussian_delay}, but for a Power-law delay model with $\gamma=0.1$ and $\kappa=0.5$, and 73 GW events in total. }
\label{data_samples_powerlaw_delay}
\end{figure}

To illustrate the procedure discussed in Sec. \ref{sec_frame}, we generate a  mock data set of LISA GW events for SMBH binary mass in the range $\[10^5 \Msun, 10^8 \Msun\]$. We assume this mass range based on two reasons. Firstly, the fundamental relationship between SMBH and the host galaxy is valid for black holes with masses larger than $10^5\Msun$, because of the lack of observations with massive black holes below $10^5\Msun$. Secondly,  LISA is unable to detect SMBH merger events with masses larger than $10^8 \Msun$ as the corresponding  frequencies are lower than the LISA frequency band.   
We sample our mock GW data following the merger rate function $\mathcal{R}( M, z | {\bf \Lambda})$, where the best-fit values are applied for ${\bm \lambda_1}$ and $ {\bm \lambda_2}$.
In particular, we consider a ``Gaussian" delay model with parameters $\alpha=-1/16$ and $\beta=\sigma=0.8\text{Gyr}$, and a Power-law delay model with parameters $\gamma=0.1$ and $\kappa=0.5$. The resulting SMBH merger rates predicted from these two delay models are shown in Fig. \ref{RBH_z_models}. 
Assuming an observation duration of four years, the total number of events in each mock data set is 78 and 73 respectively. The binary mass M and redshift z of these mock GW events are shown in Fig. \ref{data_samples_Gaussian_delay} and \ref{data_samples_powerlaw_delay}, as sampled from $\mathcal{R}( M, z | {\bf \Lambda})$. The error bar or the posterior distribution for each parameter in an individual event is estimated according to the sensitivity curve of LISA \citep{Barack:2003fp, Robson:2018ifk} and Fisher information matrix analysis using the original phenomenological waveform PhenomA \citep{Ajith_2007, Ajith_etal_2011}. We generate $10^4$ samples for $M$ and $z$ in each GW event.  


We now estimate the delay time distribution of SMBH binary mergers with these mock GW data (Fig.~\ref{data_samples_Gaussian_delay} and \ref{data_samples_powerlaw_delay}) using hierarchical Bayesian inference. 
Note that the Likelihood function defined in Eq.~(\ref{redef_likelihood})  requires a four-dimensional integration, and we are dealing with $N_{\text{det}}\times N_{\text{samples}}$ samples of GW data (where $N_{\text{samples}}$ is the number of samples of $M$ and $z$ in each GW event).
Thus for each MC sample of the hyperparameters ${\bf \Lambda}$, one needs to perform four-dimensional integration for once and two-dimensional integration for $N_{\text{det}}\times N_{\text{samples}}$ times. If one needs to generate $N_{\text{h-p}}$ samples for hyperparameters, the times of integration are at least $N_{\text{h-p}}$ and $N_{\text{det}}\times N_{\text{samples}} \times N_{\text{h-p}}$ for the four-dimensional and two-dimensional integration respectively, which is usually time-consuming. 
In addition, for the averaged Likelihood considered in Eq.~(\ref{redef_likelihood_errors}), the required times of integration becomes $N_{\text{h-p}} \times N_{\bf \lambda}$ and $N_{\text{det}}\times N_{\text{samples}} \times N_{\text{h-p}} \times N_{\bf \lambda}$ (where $N_{\bf \lambda}$ is number of samples of ${\bm \lambda_1}$ and ${\bm \lambda_2}$), which is even more computationally expensive. 
Therefore, to reduce the computational cost, we either adapt a fixed galaxy merger rate and the $M$--$M_*$ relationship by taking the best-fit value of their fitting parameters ${\bm \lambda_1}$ (Table \ref{table_galaxy_merger_rate} and  \ref{tab:Schechter_parameters}) and ${\bm \lambda_2}$ (Eq.~(\ref{M_Mstar_relation})), or optimize our numerical code to calculate the averaged likelihood.  
To be more specific, for the first method, we infer the delay-time parameters from our mock GW data using the following likelihood function:
\m \label{likelihood_parameter_estimate}
\mathcal{L}(\{{\bf d}\}|{\bm \Lambda}, {\bm {\lambda_1}}^{\text{bf}}, {\bm {\lambda_2}}^{\text{bf}}) \propto   
\prod^{N_{\text{det}}}_{i=1} \langle {P ({\bm \theta}| {\bm \Lambda}, {\bm {\lambda_1}}^{\text{bf}}, {\bm {\lambda_2}}^{\text{bf}}) \over P_{\varnothing} ({\bm \theta})} \rangle \,,
\n 
where ${\bm {\lambda_1}}^{\text{bf}}$ and ${\bm {\lambda_2}}^{\text{bf}}$ are the best-fit value for ${\bm {\lambda_1}}$ and ${\bm {\lambda_2}}$. This Likelihood has marginalised upon the rate assuming a Jeffrey’s prior $(1/N)$ for the rate function, equivalently, it makes use of the distribution information of the merger rate without considering the information of the ``observed" event number of GW events $N$. We have not found  significant differences with or without including  the N data in the Likelihood above. Rather, the uncertainties from some of the parameters in ${\bm \lambda_1}$ and $ {\bm \lambda_2}$ may result in order unity changes of $N({\bf \Lambda})$. Therefore $N({\bf \Lambda})$ is more susceptible to the model uncertainty of galaxy mergers and galaxy-SMBH relationship.
For the second case we sample the posteriors with the averaged likelihood defined in Eq.~(\ref{redef_likelihood}) and Eq.~(\ref{redef_likelihood_errors}) where the uncertainties of ${\bm {\lambda_1}}$ and ${\bm {\lambda_2}}$ are considered. The calculation of the likelihood is sped up by orders of magnitude with a Fortran code written by ourselves, and exported to Python to present the posterior samples. 

\begin{figure} 
\centering
\includegraphics[height=7cm]{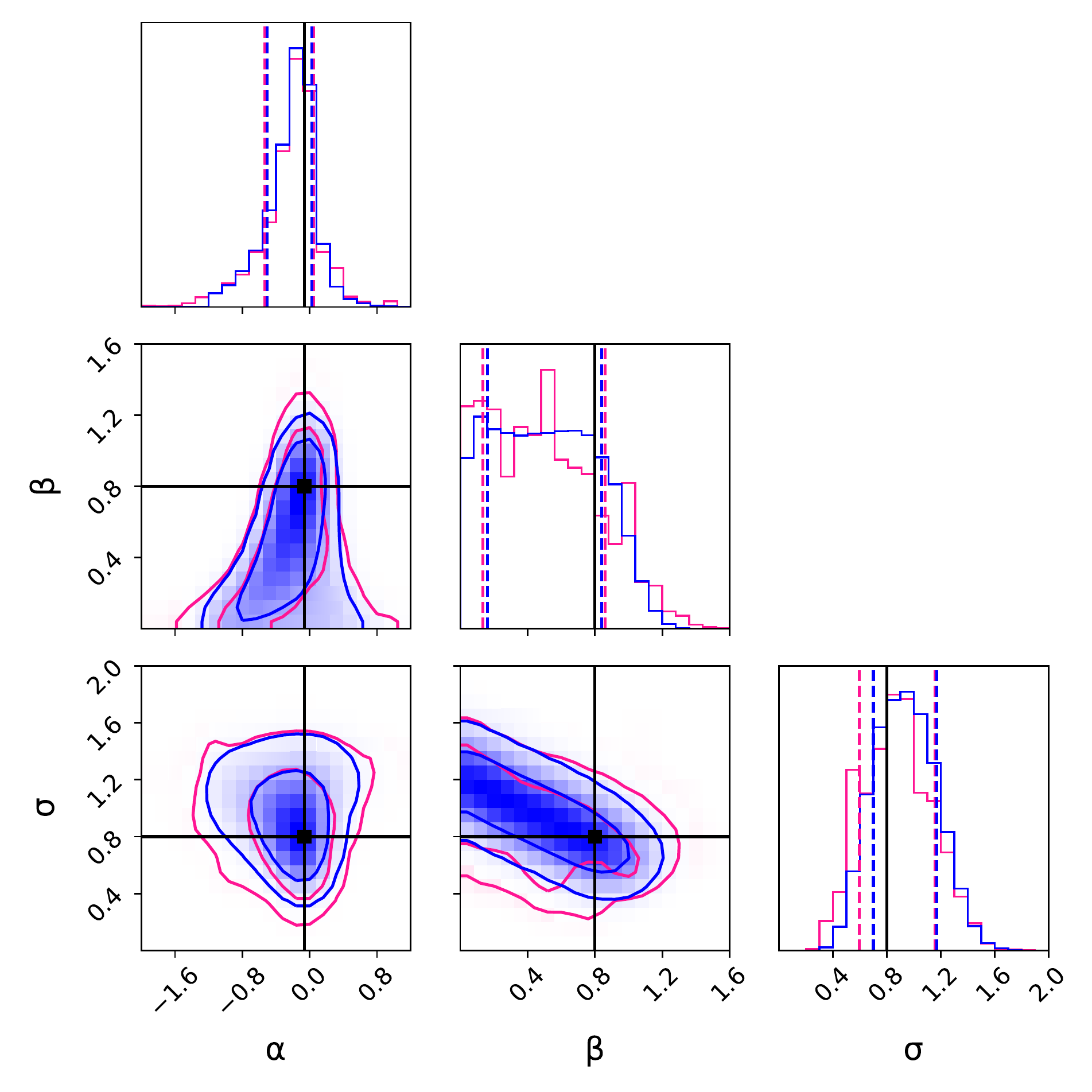}
\caption{The estimated posteriors of delay parameters in a ``Gaussian" delay distribution model from the mock GW data showed in Fig. \ref{data_samples_Gaussian_delay}. The blue lines and regions give the recovered posteriors estimated from the model assuming the best fit of fitting parameters in the galaxy merger rate and SMBH-galaxy relationship (corresponding to the likelihood function given in Eq.~\ref{likelihood_parameter_estimate}). The pink lines give the recovered posteriors estimated from the model with the uncertainties in the galaxy merger rate and SMBH-galaxy relationship included (corresponding to the averaged likelihood given in Eq.~\ref{redef_likelihood_errors}). The inner and outer circles/curves in the two-dimensional posteriors represent the $68\%$ and $95.4\%$ confidence regions. 
The injected values of the hyperparameters
are $\alpha=-1/16$, $\beta=0.8 \text{Gyr}$ and $\sigma=0.8 \text{Gyr}$, denoted by crossed black lines.
The dashed lines indicate the boundaries of the $68\%$ confidence region. 
} 
\label{delay_posterior_gaussian}
\end{figure}

\begin{figure} 
\centering
\includegraphics[height=6cm]{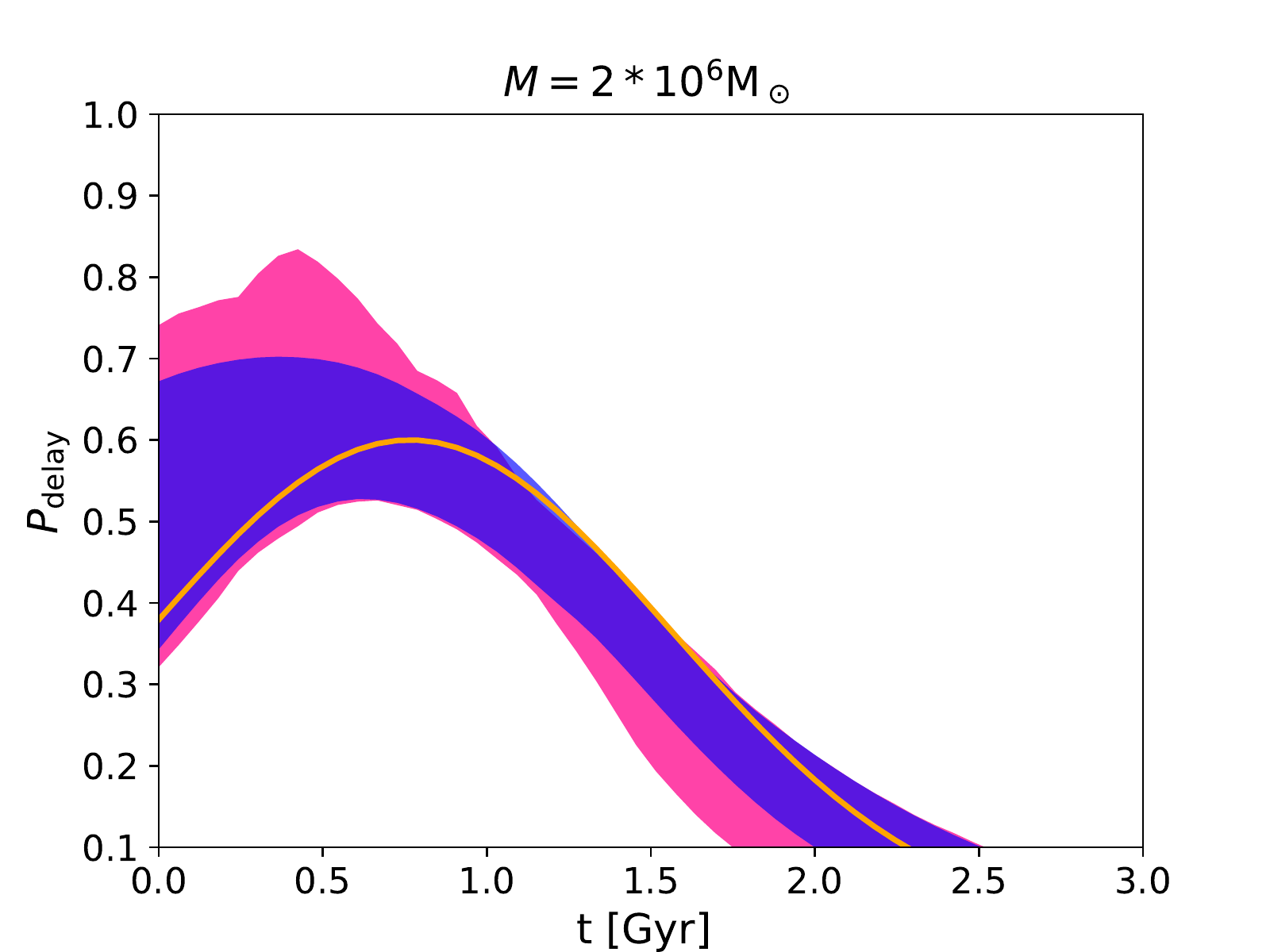}
\caption{The confidence plot of the ``Gaussian" delay distribution reconstructed according to the estimated delay parameters shown in Fig.~\ref{delay_posterior_gaussian}, using Eq.~\ref{gaussian_delay} and taking the mass of the SMBH binary to be $M=2 \times 10^6 \Msun$. The filled blue area is the one-sigma confidence region for the model considered in the blue lines of Fig.~\ref{delay_posterior_gaussian}. The filled pink area is the one-sigma confidence region for the model considered in the pink lines of Fig.~\ref{delay_posterior_gaussian}. The orange line is the given truth delay distribution according to Eq.~(\ref{gaussian_delay}). 
}
\label{confidence_delay_gaussian}
\end{figure}

\begin{figure} 
\centering
\includegraphics[height=6cm]{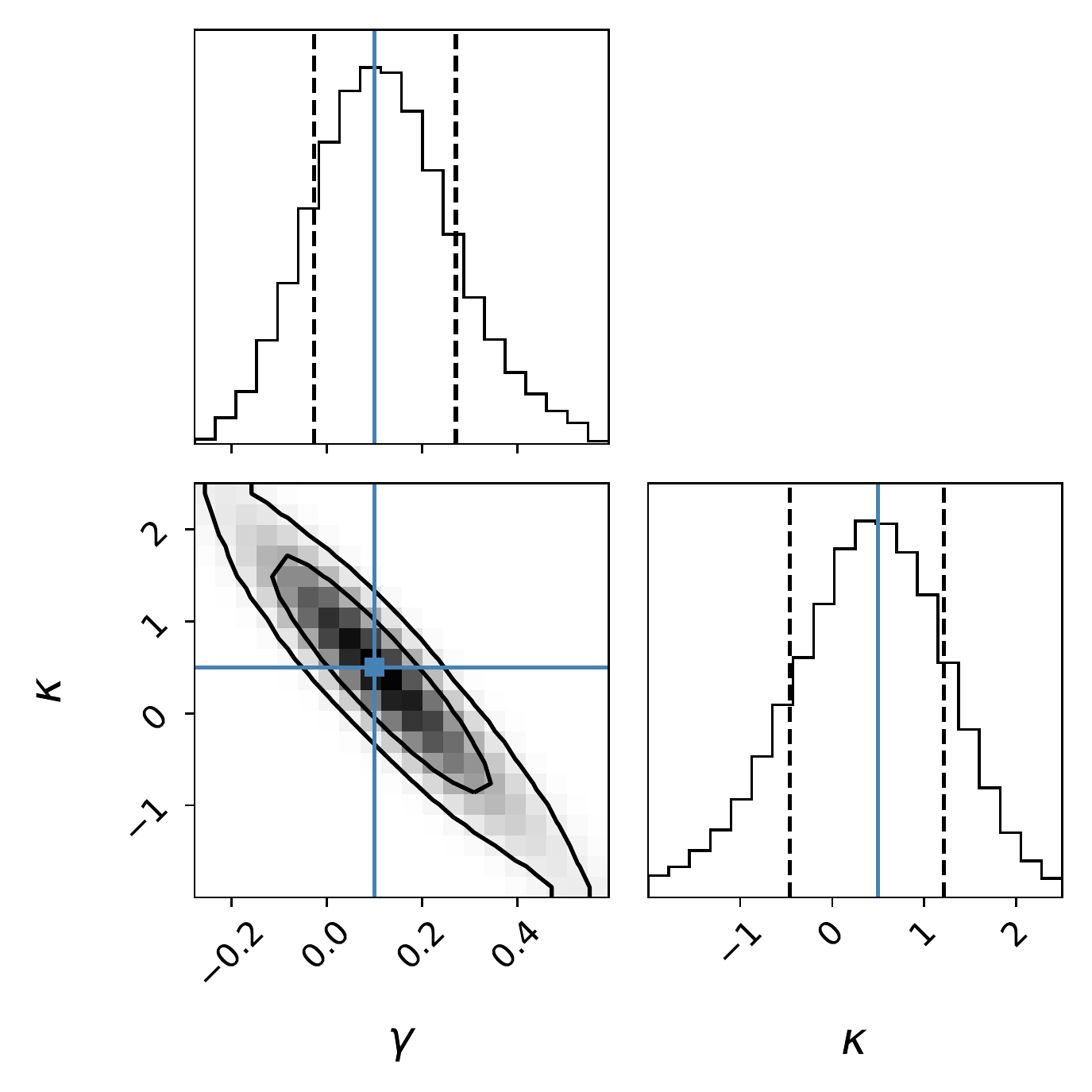}
\caption{The estimated posteriors of delay parameters in a Power-law delay model from the mock GW data showed in Fig.  \ref{data_samples_powerlaw_delay}. The posterior is estimated assuming the best-fit value of the fitting parameters. 
The truth values of the hyperparameters are $\gamma=0.1$ and $\kappa=0.5$, denoted by crossed blue lines. The black dashed lines indicate the boundaries of the $68\%$ confidence region. }
\label{delay_posterior_powerlaw}
\end{figure}

\begin{figure} 
\centering
\includegraphics[height=6cm]{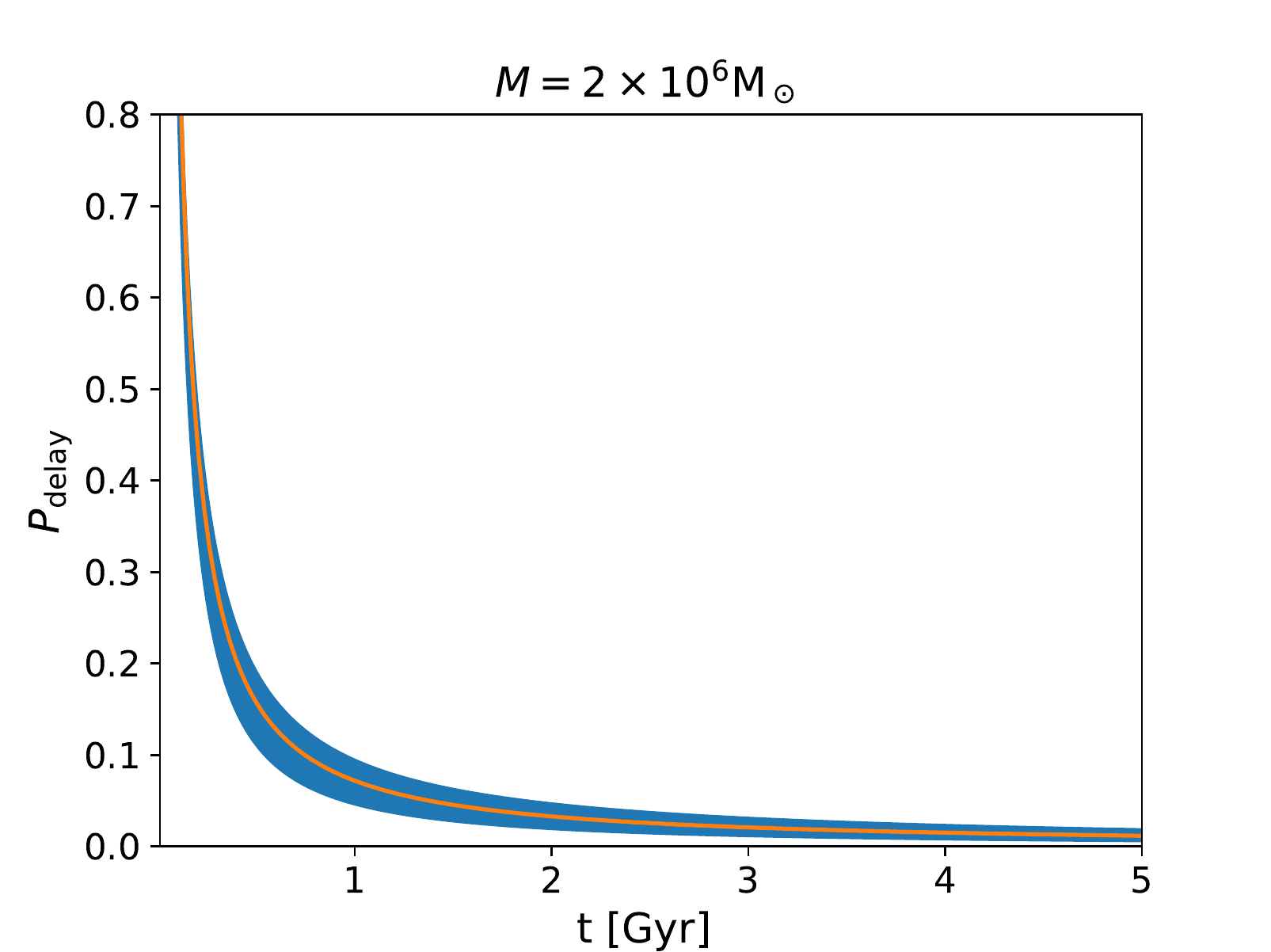}
\caption{The confidence plot of the Power-law delay time distribution reconstructed according to the estimated delay parameters shown in Fig. ~\ref{delay_posterior_powerlaw}, using Eq. \ref{powerlaw_delay} and taking the mass of the SMBH binary to be $M=2 \times 10^6 \Msun$. The filled blue area is the one sigma confidence region, and the orange line is the given truth delay distribution according to Eq.~(\ref{powerlaw_delay}). 
 }
\label{confidence_delay_powerlaw}
\end{figure}

As illustrative examples, the posteriors of delay parameters for the ``Gaussian" delay and Power-law delay models are shown in Fig. \ref{delay_posterior_gaussian} and \ref{delay_posterior_powerlaw} respectively. Based on the posteriors of the hyperparameters, the confidence plot of the delay functions can be displayed for various SMBH masses. For example, the recovered delay time distribution for SMBH binary with mass $M=2 \times 10^{6}\Msun$ are shown in Fig. \ref{confidence_delay_gaussian} and \ref{confidence_delay_powerlaw}.  
In the ``Gaussian" delay model (Fig. \ref{delay_posterior_gaussian} and Fig. \ref{confidence_delay_gaussian}), we use two different likelihood functions to perform the estimations. One is the averaged likelihood function defined in Eq.~(\ref{redef_likelihood_errors}) which takes into account the uncertainties from the galaxy merger rate model and the SMBH-galaxy relationship. Another one is the likelihood given in Eq.~(\ref{likelihood_parameter_estimate}) which simply assuming the best fit of these model and relationship. Comparing the results estimated from the averaged likelihood (the pink lines in Fig.~\ref{delay_posterior_gaussian} and the pink-shaded region in Fig.~\ref{confidence_delay_gaussian}) with the results estimated from the likelihood assuming best fits (the blue lines in Fig. \ref{delay_posterior_gaussian} and blue-shaded region in Fig. \ref{confidence_delay_gaussian}), we see that the uncertainties from the galaxy merger rate and fundamental relationship we assumed here only mildly widen the errors of the recovered hyperparameters (Fig.~\ref{delay_posterior_gaussian}) and therefore mildly widen the reconstructed delay time distribution (Fig.~\ref{confidence_delay_gaussian}). 
Fig. \ref{confidence_delay_gaussian} shows that the delay time distribution in both two cases is better constrained at the larger delay time region than the smaller delay time region. This is due to the fact that the larger delay time induces a larger shift in the merger distribution over redshift, while the smaller delay time induces a smaller shift in redshift, thus leading to merger distributions being less distinguishable from each other.
There is also a degeneracy between $\beta$ and $\sigma$ in the ``Gaussian" delay model as shown in Fig. \ref{delay_posterior_gaussian}. This is mainly due to the fact that the delay time scale is simultaneously determined by $\beta$ and $\sigma$ (Eq.~(\ref{gaussian_delay})). 

In the Power-law delay model, we only consider the case assuming the best-fit of fitting parameters, i.e., taking ${\bm {\lambda_1}}^{\text{bf}}$ and ${\bm {\lambda_2}}^{\text{bf}}$ in the likelihood. We have tried the posterior samples with the averaged likelihood which includes the uncertainties of ${\bm {\lambda_{1,2}}}$, while found the result hard to converge. This is because the SMBH merger rate  modeled with the Power-law delay model is much more sensitive to the choice of ${\bm {\lambda_{1,2}}}$, thus needs higher precision to do the integration (discussed in section \ref{estimate_delay_posteriors}) in likelihood and more samples of ${\bm {\lambda_{1,2}}}$ to do the average of likelihood which is much more computationally expensive. Nevertheless, by doing the samples using the likelihood with the best fits, one could test in principle, whether the method works for the specific delay model. 
From the results illustrated in Fig. \ref{delay_posterior_powerlaw} we find that thanks to fewer hyperparameters in Power-law delay model compared to the ``Gaussian" delay model, the delay parameters $\gamma$ and $\kappa$ are better constrained, with the true value of both parameters settled in the center of the posterior distribution and the shape of the distribution is close to Gaussian. And thus the delay distribution is also better constrained as shown in Fig.\ref{confidence_delay_powerlaw}. It is clear from these figures that the delay parameters and the delay time distributions are both properly recovered with the inference method we proposed here. 

So far, we have shown that the delay time distribution of SMBH binary coalescence can in principle be recovered using hierarchical Bayesian inference from LISA GW data. This highlights the main point of this work, that the delay time distribution as an observable may be measured with GW events. With our Fortran code developed for the likelihood calculation, we have shown that it is possible to include the model uncertainties in the inference procedure for the ``Gaussian" delay model. Similar task is challenging for the ``Power-law" delay model because of the poor convergence.
In reality, according to the predictions of the delay time distribution from previous works \citep[see, e.g.,] []{Chen_2020}, the proposed ``Gaussian" delay model is probably more scientifically sound  than the Power-law distribution. 
On the other hand, it is also worth mentioning that the current understanding of the galaxy merger rate and the SMBH-galaxy relationship (especially at high redshift) is currently subject to large uncertainties either from observations or theoretical predictions.  
 It is expected that future observations of galaxy mergers together with large-scale cosmological simulations will improve the accuracy of these relations, which is crucial to minimize the systematic error in the measurement of the delay time distribution. 


%

\section{Comparing different delay time distribution models}
\label{compare_delay_model}
The recovered delay time distribution encodes important information about the evolution mechanism of SMBH binaries. As discussed in Sec. \ref{sec_delay_models}, different evolutionary scenarios predict different delay times, so that the measured delay time distribution may be used to test the predictions of different models. 
In this section, we perform the comparison between models based on the ``observed" time delay distribution using the Bayesian Model Selection Method. This method compares the statistical significance of different delay models by calculating the Bayesian Odds ratio, which is a statistical quantity that measures the relative degree being favored by data for two competing models. The standard range of Odds ratio in determining the strength of evidence is listed in Table \ref{tab:Bayes_factor}. 

The comparison can be made be between  phenomenological delay models which are parameterized with hyperparameters, such as the ``Gaussian" delay and the Power-law delay models; or it can be made between delay time models with fixed delay time distributions, such as those predicted from dedicated simulations or analytical models. 
The Odds ratio for the later is given by
 \m \label{Odds_ratio}
 \mathcal{O}={ P({\bf \Lambda}_A|\{ {\bf d} \}) \over P({\bf \Lambda}_B|\{ {\bf d} \})} = { P(\{ {\bf d \} }| {\bf \Lambda}_A ) \over P(\{ {\bf d} \} | {\bf \Lambda}_B )} { P({\bf \Lambda}_A ) \over P( {\bf \Lambda}_B )}
 = \mathcal{B} { P({\bf \Lambda}_A ) \over P( {\bf \Lambda}_B )} \,,
\n 
where ${\bf \Lambda}_A$ and ${\bf \Lambda}_B$ represent the given hyperparameters for delay model A and B, and the Bayes factor $\mathcal{B}$ is the ratio of the Bayesian evidence between the two models. Here, we assume $ P({\bf \Lambda}_A ) = P( {\bf \Lambda}_B )$, so that the Odds ratio $\mathcal
{O}$ equals $\mathcal{B}$. 
The Odds ratio for the comparison between parametrized phenomenological delay models is given by 
\m \label{Odds_ratio}
 \mathcal{O}={P(A)\over P(B)}{ \int P({\bf \Lambda}_A|\{ {\bf d} \}) \ \text{d}{\bf \Lambda}_A  \over \int P({\bf \Lambda}_B|\{ {\bf d} \}) \text{d}{\bf \Lambda}_B } = \mathcal{B} { P(A) \over P(B)} \,,
\n 
where $P(A)$ and $P(B)$ are the prior of model $A$ and $B$.
The evidence is given by the integration of the posterior distribution as inferred from the hierarchical Bayesian approach.  
As discussed in section \ref{estimate_delay_posteriors},  hierarchical Bayesian inference with the averaged likelihood for the Power-law delay model has the problem of non-efficient numerical convergence, so is the evidence for this delay model. Thus there is a challenge to perform the comparison between the phenomenological ``Gaussian" and Power-law delay models using the averaged Likelihood, though the ``Gaussian" delay model does converge efficiently. 

In order to illustrate the ability of Bayesian model selection, here, we compare delay models assuming fixed delay time distribution. As a proof of concept, we compare different delay models distinguished by different model parameters in the phenomenological ``Gaussian" delay distribution proposed in Eq.~(\ref{gaussian_delay}). As  discussed in \ref{sec_delay_models}, the stellar and gas-rich dynamical scenarios predict delay time with different scale and mass dependence, the simplified expression is roughly proportional to $\beta  M^\alpha$. The ``Gaussian" delay proposed in Eq.~(\ref{gaussian_delay}), with $\beta$ characterizes the delay time scale, $\alpha$ parameterizes the dependence on the mass of binary and $\sigma$ characterizes the dispersion of the delay time due to the statistical property of the environment the binary resides in, describes various delay distributions produced by different binary dynamical scenarios. The specific models we assume for model comparison are listed in Table \ref{tab:gaussian_delay_models}.  

\begin{figure} 
\centering
\includegraphics[height=7cm]{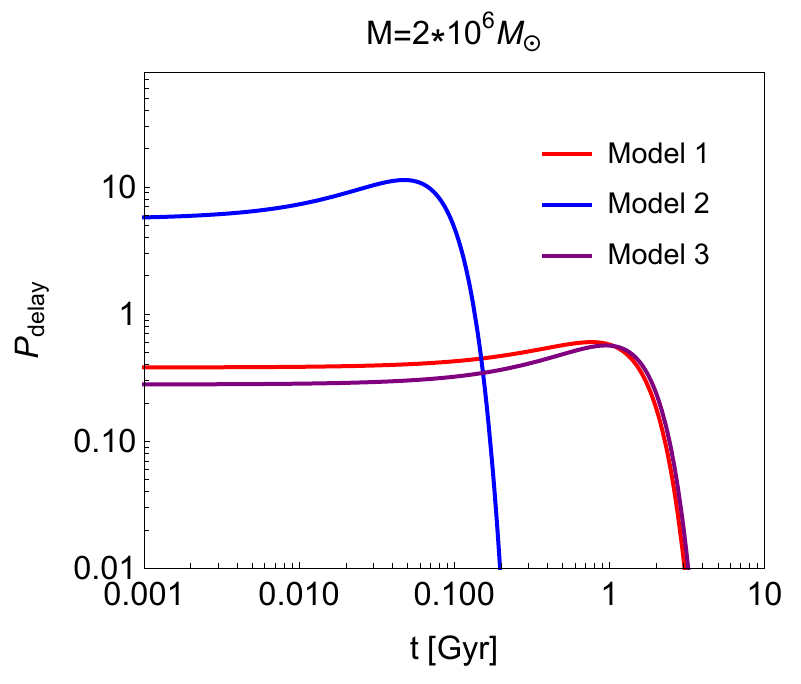}
\caption{The delay time distribution for the three delay models considered in Table \ref{tab:gaussian_delay_models} parameterized as a ``Gaussian" distribution given in Eq.~(\ref{gaussian_delay}), taken the mass of the SMBH binary to be $2 \times10^6\Msun$.
}
\label{delay_gaussian_examples}
\end{figure}

\begin{figure} 
\centering
\includegraphics[height=7cm]{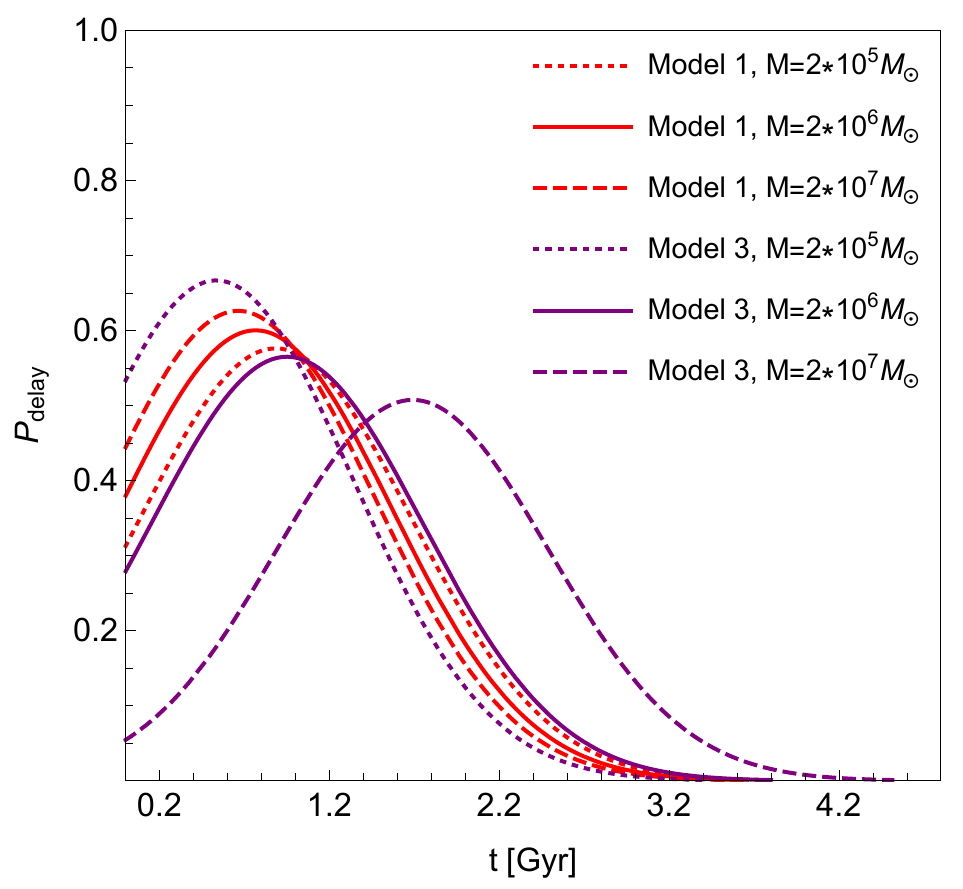}
\caption{The delay time distribution referred to model 1 and model 3 in Table \ref{tab:gaussian_delay_models} for different masses of SMBH binary. 
}
\label{delay_gaussian_various_mass}
\end{figure}

The corresponding delay time distributions for the three models we take in Table \ref{tab:gaussian_delay_models} are plotted in Fig. \ref{delay_gaussian_examples}, assuming the binary mass to be $M=2\times 10^6 \Msun$. The blue line (model 1) suggests a delay time of order $10\text{Myrs}$, and the red (Model 1) and purple (Model 3) lines correspond to a  delay time of  $\mathcal{O}(\text{Gyr})$. For a intuitive sense, the resulting marginalized SMBH binary merger rate predicted by these three delay models assuming the best fit values for parameters ${\bm \lambda_{1,2}}$ are shown previously in Fig. \ref{RBH_z_models}.  
Fig. \ref{delay_gaussian_various_mass} illustrates the mass dependence of delay time distribution of Model 1 and Model 3, assuming the binary mass to be $M=2\times 10^5 \Msun$, $2\times 10^6 \Msun$, and $2\times 10^7 \Msun$ respectively. Although for $M=2\times 10^6 \Msun$, Model 1 and Model 3 share a similar delay time scale, for larger mass such as the $M=2\times 10^7 \Msun$ case they differ significantly, as shown in Fig. \ref{delay_gaussian_various_mass}. The delay time distribution in Model 3 is more sensitive to mass than Model 1, i.e., Model 3 predicts longer delay times for larger masses, while Model 1 has a weaker and opposite trend. As a result, Model 1 and Model 3 actually predict different joint distribution in merger rate over redshift and mass.  
The mock GW data we use for model comparison is the same to the one given in Fig. \ref{data_samples_Gaussian_delay}, which is generated from the merger rate assuming Model 1 for the delay time model and taking the best-fit values for the parameters ${\bm \lambda_{1,2}}$.


%
\begin{table} 
{
\centering
\caption{The delay parameters for different ``Gaussian" delay models, and the result of Odds ratio between models.}
\label{tab:gaussian_delay_models}
\begin{tabular}{  c | c  c  c | c   }
\hline
Model & $\alpha$ & $\beta$ ($\text{Gyr}$) &  
$\sigma$ ($\text{Gyr}$)  & Odds ratio  \\ 
\hline
 1 & -1/16 & 0.8 & 0.8 &  \\[0.15cm]
 2 & 1/4 & 0.04 & 0.04 & $\sim 10^7$ \\[0.15cm]
 3 & 1/4 &  0.8 & 0.8  &  327 or 524 \\[0.15cm]
\hline
\end{tabular}
\par
} 
\begin{tablenotes}
     \item {\bf Notes}: 
     Model 1 represents the delay time model predicted in a stellar dynamical scenario described in Sec. \ref{delay_stellar}, model 2 represents the case of a gas-rich dynamical scenario, as discussed in Sec. \ref{gas_rich_scenario}. Model 3 represents delay time model either from a stellar scenario with different mass dependence or the gas-rich scenario but with larger delay timescale. The last column lists the Odds ratio of Model 1 to Model 2 and Model 1 to Model 3 calculated from our averaged Likelihood function. 
\end{tablenotes}
\end{table}
\begin{table} 
{
\centering
\caption{The interpretation of the Odds ratio (or Bayes factor) in determining the strength of evidence the model is favored \citep[see][]{Kass_Raftery_1995}.}
\label{tab:Bayes_factor}
\begin{tabular}{  c  c   }
\hline
$\mathcal{O}$ & Strength of evidence  \\ 
\hline
 $<1$ & Negative  \\[0.15cm]
 1-3.2 & Not worth more than a bare mention  \\[0.15cm]
 3.2-10 & Weak   \\[0.15cm]
 10-100 & Strong   \\[0.15cm]
$>100$ & Very strong/Decisive   \\[0.15cm]
\hline
\end{tabular}
\par
} 
\end{table}

We use the averaged Likelihood defined in Eq.~(\ref{redef_likelihood_errors}) to calculate the Bayes factor and Odds ratio in Eq.~(\ref{Odds_ratio}), in which the modeling uncertainties from the fitting parameters ${\bm \lambda_1}$ and ${\bm \lambda_2}$ in galaxy merger rate and $M$--$M_*$ relationship are taken into account. Specifically, we generate samples for each  parameter in ${\bm \lambda_1}$ (${\bm \lambda_2}$ is treated in the way described in section \ref{sec_frame}) from their PDFs. Each sample of ${\bm \lambda_{1,2}}$ gives one Likelihood $\mathcal{L}(\{{\bf d}\}|{\bm \Lambda}, {\bm \lambda_1}, {\bm \lambda_2})$, and the summation over these Likelihood functions leads to the averaged Likelihood $\bar{\mathcal{L}}(\{{\bf d}\}|{\bm \Lambda})$. We use two different definitions of Likelihood for comparison, one is given by
\m \label{redef_likelihood1}
\mathcal{L}(\{{\bf d}\}|{\bm \Lambda}, {\bm \lambda}_1, {\bm \lambda}_2 ) &\propto&  
{N({\bf \Lambda}, {\bm \lambda}_1, {\bm \lambda}_2)}^{N_{\text{det}}} e^{- N ({\bf \Lambda}, {\bm \lambda}_1, {\bm \lambda}_2) } \\ \nonumber 
&&
\prod^{N_{\text{det}}}_{i=1} \langle {P ({\bm \theta}| {\bm \Lambda}, {\bm \lambda}_1, {\bm \lambda}_2) \over P_{\varnothing} ({\bm \theta})} \rangle
\n
where the information of the number of GW events $N$ is included, 
and another one is given by 
\m \label{redef_likelihood2}
\mathcal{L}(\{{\bf d}\}|{\bm \Lambda}, {\bm \lambda}_1, {\bm \lambda}_2 ) \propto 
\prod^{N_{\text{det}}}_{i=1} \langle {P ({\bm \theta}| {\bm \Lambda}, {\bm \lambda}_1, {\bm \lambda}_2) \over P_{\varnothing} ({\bm \theta})} \rangle
\n
where the information of number of events (or the merger rate amplitude) is neglected.  
 
The Odds ratio for the comparison between Model 1 and  Model 2 is $\mathcal{O}(10^7)$ (with and without considering the factor of total number of events in the Likelihood function), indicating a decisive evidence favoring Model 1 over Model 2. This result is consistent with our expectation since the the mock GW data is sampled from Model 1 (though assuming best-fit values for parameters ${\bm \lambda_{1,2}}$), which predicts rather different scale of delay time as compared to Model 2 as shown in Fig. \ref{delay_gaussian_examples}. On the other hand, the Odds ratio for Model 1 compared to Model 3 is approximately 327 (without including the total event number in the Likelihood function) and 524 (total event number included ) respectively, still indicating a decisive evidence supporting Model 1 over Model 2. This is also consistent with our expectation as Model 1 and 3 predict different mass dependence of delay distribution (Fig. \ref{delay_gaussian_various_mass}) and thus result to different joint distribution over $M$ and $z$ in the merger rate. Including the information about the total number of merger events in the Likelihood function  further increases the evidence as expected. The delay time difference between Model 1 and Model 3 is no more than a few times (Fig. \ref{delay_gaussian_various_mass}), while it is a few orders of magnitude different between Model 1 and Model 2 (Fig. \ref{delay_gaussian_examples}). That is the reason why the Odds ratio between them also differs by orders. Nevertheless, for the specific mock data set we have used, it is possible to statistically distinguish different models based on data. This provides a promising aspect for testing  delay time models using their theoretical predictions and will promotes our understanding of SMBH/SMBH binary evolution histories.  

\section{Conclusion and discussion}
\label{conclusion}

In this work we have discussed how to apply a set of SMBH bianry merger events detected by space-borne GW detectors to study the evolution mechanisms of SMBH binaries, through the measurement of delay times. In fact, the delay time between the galaxy merger and various stages of SMBH binary evolution may all be potential targets of opportunities, but it is generally difficult to address the selection effects in electromagnetic observations in order to obtain faithful information about the SMBH binary population and distribution. Thanks to the fact that SMBH binary mergers are the energetically loudest events in the universe, mergers within the right mass range are likely to be all identified by the GW detectors such as LISA. Therefore measuring the delay time till merger through GWs seems to be the most promising way to test different binary evolution models.  

Such tests are illustrated in Sec.\ref{estimate_delay_distri} and Sec.\ref{compare_delay_model} with Mock GW data ($\sim 70$--$80$ events) generated assuming an observation period of four years. The mass and redshift measurement uncertainties are estimated with Fisher Information Matrix, using the sensitive curve of LISA \citep{Barack:2003fp, Robson:2018ifk} and the phenomenological waveform \citep{Ajith_2007, Ajith_etal_2011}. The population model of SMBH binary mergers is constructed with galaxy merger rate together with the intrinsic relationship between SMBH binary mass $M$ and descendant galaxy stellar mass $M_*$.  
The delay time distribution is inferred via the framework of hierarchical Bayesian inference. Using the likelihood including (without including) the uncertainties from the galaxy merger rate (from an Illustris simulation by \cite{Vicente2015} and observational results of Schechter function, see Tables \ref{table_galaxy_merger_rate} --\ref{tab:Schechter_parameters}) and the SMBH-galaxy relationship \citep[Eq.~(\ref{M_Mstar_relation}),][]{Kormendy:2013dxa}, our results Fig. \ref{delay_posterior_gaussian}--\ref{confidence_delay_gaussian} (Fig.~\ref{delay_posterior_gaussian}--\ref{confidence_delay_powerlaw}) show that the delay time distribution can be properly recovered within the uncertainties. 
To compare different delay models, the method of Bayesian model selection is used to quantify the relative faithfulness of different models based on data. The result (Table \ref{tab:gaussian_delay_models}) shows that at least for the models discussed in Sec.\ref{compare_delay_model}, the statistical evidence to distinguish different delay models is significant. Of course, one possible caveat of this analysis -  the systematic error of the galaxy merger rate and $M$--$M_*$ relationship, is only partially addressed by marginalizing over uncertainties of modeling parameters. This part ought to be updated with future observations of galaxy merger rate via galaxy surveys, and more precise and consistent cosmological simulations.  

We have only considered a few delay-time models motivated by simple physical driving mechanisms. 
In reality, the formation and evolution of SMBHs or SMBH binaries are influenced by more complex astrophysical conditions, such as the impact of seeding models of SMBH formations \citep{Klein:2015hvg, Toubiana:2021iuw}, the role of mass accretion for the individuals in SMBH binaries \citep{Callegari_2011}, SMBH binary mergers in multiple systems \citep{Hoffman_Loeb_2007, Bonetti:2018tpf, Bonetti:2017lnj}, and other impacts \citep{Barausse_2020}.  
In the future, more realistic delay time models  from simulations that include various  astrophysical processes, such as the processes mentioned above, should be considered when performing delay model selection. In addition to the model comparison between different ``Gaussian" delay models in section 4,  a comparison between a ``Gaussian" delay model and a Power-law delay taking GW samples predicted from the latter and assuming best-fit parameters ${\bm \lambda_{1,2}}$ is straightforward, and the result   strongly supports the Power-law delay just as expected.  
The SMBH binaries in the mass region ($10^5\Msun-10^8 \Msun$) we consider in this work is related to the heavy seed models. It will be interesting to extend the data analysis to MBH binaries with smaller masses at high redshift which includes the information of light seeds mergers.  
All these effects should in principle contribute to the final distribution of SMBH binary mergers, and should be considered in a systematic framework of population analysis. In addition, as we have only discussed events detected by LISA, it will be interesting and straightforward to extend the framework discussed here to also include possible PTA events.

\section*{Acknowledgements}
We acknowledge the use of the HPC Cluster at Perimeter Institute and the HPC Cluster of the National Supercomputing Center in Beijing. This work makes use of the open-sourced python package emcee \citep{Foreman2013emcee}. 
We thank Reed Essick, Youjun Lu and Zhenwei Lyu for useful discussions. We also thank referees for constructive comments. 
Y. F. is supported in part by the National Science Foundation
of China (NSFC) Grant No. 11721303, the fellowship of China Postdoctoral Science Foundation No. 2021M690228, and in part by Perimeter Institute for Theoretical Physics. 
HY is supported by the Natural Sciences and Engineering Research
Council of Canada and in part by Perimeter Institute for Theoretical Physics. 
Research at Perimeter Institute is supported by the Government of Canada through the Department of Innovation, Science, and
Economic Development, and by the Province of Ontario through the Ministry of Colleges and Universities. 

\section*{Data availability}
We provide our Fortran and Python code and the mock data of LISA GW events in our population analysis in \cite{fang_SMBHB_delay}. We have made use of the Fortran code by \cite{Zhang_Jin_1996} for the evaluation of the hypergeometric function, and we made a change such that the argument could include the case of $\text{abs}(z) > 1$.  

\appendix
\section{Galaxy merger rate model}
\label{sec.app}

In the appendix, we present the galaxy merger rate per galaxy and stellar mass function  used in the main text in Table \ref{table_galaxy_merger_rate} and \ref{tab:Schechter_parameters}. 
\begin{table*}
		\centering
  		\caption{the fitting function and best-fitting parameters for the galaxy-galaxy merger rate per galaxy taken from \citet{Vicente2015}, here $M_*$ and $\mu$ correspond to stellar mass and stellar mass ratio. }
		\begin{tabular}{c  c  c}
			\hline
			Definition & $\frac{{\rm d}N}{{\rm d}\mu \, {\rm d}t} (M_{*}, \mu, z)$ \\
			Units & Gyr$^{-1}$ \\
			\hline
			\parbox{3cm}{Fitting function} &
			\parbox{8.5cm}{\centering $A(z) \, \left(\frac{M_{*}}{10^{10}\Msun}\right)^{\alpha(z)} \left[1 +  \left(\frac{M_{*}}{M_0}\right)^{\delta(z)}\right] \mu^{\beta (z) + \gamma \log_{10}\left(\frac{M_{*}}{10^{10}\Msun}\right)}$, \\
			                               where \\
			                               $A(z) = A_0 (1+z)^{\eta}$, \\
			                               $\alpha(z) = \alpha_0 (1+z)^{\alpha_1}$, \\
			                               $\beta(z) = \beta_0 (1+z)^{\beta_1}$, \\
			                               $\delta(z) = \delta_0 (1+z)^{\delta_1}$, \\
			                               and $M_0 = 2 \times 10^{11} \, \Msun$ is fixed.} \\
			\parbox{2cm}{\centering
				$\log_{10}(A_0/{\rm Gyr^{-1}})$ \\
				$\eta$ \\
				$\alpha_0$ \\
				$\alpha_1$ \\
				$\beta_0$ \\
				$\beta_1$ \\
				$\gamma$ \\
				$\delta_0$ \\
				$\delta_1$} &
			\parbox{4cm}{\flushright
				$-2.2287 \pm 0.0045$ \\
				$2.4644 \pm 0.0128$ \\
				$0.2241 \pm 0.0038$ \\
				$-1.1759 \pm 0.0316$ \\
				$-1.2595 \pm 0.0026$ \\
				$0.0611 \pm 0.0021$ \\
				$-0.0477 \pm 0.0013$ \\
				$0.7668 \pm 0.0202$ \\
				$-0.4695 \pm 0.0440$} \hspace{2cm} \\

			\hline
		\end{tabular} \label{table_galaxy_merger_rate}
	\end{table*} 


\begin{table*}
\caption{The best-fitting double (lower section) and single (upper section) Schechter function parameters.}
\begin{tabular}{  c  c  c  c  c  c  c  }
\hline
Redshift & $\mathcal{M}^{\star}$ & $\log_{10} (\phi_{1}^{\star})$ &  
$\alpha_{1}$ &  $\log_{10}(\phi_{2}^{\star})$ & $\alpha_{2}$ & ref.  \\ 
\hline
$0 \leq z < 0.25$      & 10.66 $\pm$ 0.05 & $-$2.40 $^{+0.04}_{-0.04}$ &  $-$0.35 $\pm$ 0.18 & $-$3.10 $^{+0.11}_{-0.15}$ & $-$1.47 $\pm$ 0.05 &  \\[0.15cm]
$0.25 \leq z < 0.75$ & 10.80 $\pm$ 0.06 & $-$2.77 $^{+0.06}_{-0.07}$ &  $-$0.61 $\pm$ 0.23 & $-$3.26 $^{+0.12}_{-0.17}$ & $-$1.52 $\pm$ 0.05 &   \\[0.15cm]
$0.75 \leq z < 1.25$ & 10.72 $\pm$ 0.07 & $-$2.80 $^{+0.07}_{-0.09}$ &  $-$0.46 $\pm$ 0.34 & $-$3.26 $^{+0.15}_{-0.23}$ & $-$1.53 $\pm$ 0.07 &  \cite{McLeod2021},    \\[0.15cm]
$1.25 \leq z < 1.75$ & 10.72 $\pm$ 0.05 & $-$2.94 $^{+0.04}_{-0.05}$ &  $-$0.55 $\pm$ 0.22 & $-$3.54 $^{+0.14}_{-0.22}$ & $-$1.65 $\pm$ 0.07 & \cite{Baldry2012}   \\[0.15cm]
$1.75 \leq z < 2.25$ & 10.77 $\pm$ 0.06 & $-$3.18 $^{+0.07}_{-0.08}$ &  $-$0.68 $\pm$ 0.29 & $-$3.84 $^{+0.22}_{-0.46}$ & $-$1.73 $\pm$ 0.12 &     \\[0.15cm]
$2.25 \leq z < 2.75$ & 10.77 $\pm$ 0.10 & $-$3.39 $^{+0.09}_{-0.11}$ &  $-$0.62 $\pm$ 0.50 & $-$3.78 $^{+0.23}_{-0.50}$ & $-$1.74 $\pm$ 0.13 &    \\[0.15cm]
$2.75 \leq z < 3.75$ & 10.84 $\pm$ 0.18 &$-$4.30 $^{+0.23}_{-0.52}$ & $-$0.00 $\pm$ 1.03 & $-$3.94 $^{+0.20}_{-0.37}$   & $-$1.79 $\pm$ 0.09 &    \\[0.15cm]
\hline
Redshift & $\mathcal{M}^{\star}$ & $\log_{10} (\phi^{\star})$ &  
$\alpha$ &  &  & ref.  \\ 
\hline
$3.75<z<4.5$ & 10.96 $\pm$ 0.13 & $-$3.94 $\pm$ 0.16 &  $-$1.63 $\pm$ 0.05 &  &  &  \\[0.15cm]
$4.5<z<5.5$ & 10.78 $\pm$ 0.23 & $-$4.18 $\pm$ 0.29 & $-$1.63 $\pm$ 0.09 &  &  & \cite{Grazian2015AA} \\[0.15cm]
$5.5<z<6.5$ & 10.49 $\pm$ 0.32 & $-$4.16 $\pm$ 0.47 & $-$1.55 $\pm$ 0.19 &  &  & \\[0.15cm]
\hline
Redshift & $\mathcal{M}^{\star}$ & $\log_{10} (\phi^{\star})$ &  
$\alpha$ &  &  & ref.  \\ 
\hline
$6.5 \leq z < 7.5$ & $10.04^{+0.15}_{-0.13}$ & $-4.14^{+0.19}_{-0.23}$ & $-1.73^{+0.08}_{-0.08}$ &  &  & \\[0.15cm]
$7.5 \leq z < 8.5$ & $ 9.98^{+0.44}_{-0.24}$ & $-4.69^{+0.40}_{-0.72}$ &  $-1.82^{+0.20}_{-0.21}$ &  &  & \cite{Stefanon_2021}  \\[0.15cm]
$8.5 \leq z < 9.5$ & $ 9.50$ [fixed] & $-5.12^{+0.10}_{-0.13}$ & $-2.00$ [fixed]  & &  &  \\[0.15cm]
$9.5 \leq z < 10.5$ & $ 9.50$ [fixed] & $-6.13^{+0.19}_{-0.36}$ &  $-2.00$ [fixed] & &  &  \\[0.15cm]
\hline
\end{tabular}
\label{tab:Schechter_parameters}
\end{table*}

\bibliographystyle{mnras}
\bibliography{reference}

\begin{thebibliography}{}
\makeatletter
\relax
\def\mn@urlcharsother{\let\do\@makeother \do\$\do\&\do\#\do\^\do\_\do\%\do\~}
\def\mn@doi{\begingroup\mn@urlcharsother \@ifnextchar [ {\mn@doi@}
  {\mn@doi@[]}}
\def\mn@doi@[#1]#2{\def\@tempa{#1}\ifx\@tempa\@empty \href
  {http://dx.doi.org/#2} {doi:#2}\else \href {http://dx.doi.org/#2} {#1}\fi
  \endgroup}
\def\mn@eprint#1#2{\mn@eprint@#1:#2::\@nil}
\def\mn@eprint@arXiv#1{\href {http://arxiv.org/abs/#1} {{\tt arXiv:#1}}}
\def\mn@eprint@dblp#1{\href {http://dblp.uni-trier.de/rec/bibtex/#1.xml}
  {dblp:#1}}
\def\mn@eprint@#1:#2:#3:#4\@nil{\def\@tempa {#1}\def\@tempb {#2}\def\@tempc
  {#3}\ifx \@tempc \@empty \let \@tempc \@tempb \let \@tempb \@tempa \fi \ifx
  \@tempb \@empty \def\@tempb {arXiv}\fi \@ifundefined
  {mn@eprint@\@tempb}{\@tempb:\@tempc}{\expandafter \expandafter \csname
  mn@eprint@\@tempb\endcsname \expandafter{\@tempc}}}

\bibitem[\protect\citeauthoryear{Abbott et~al.}{Abbott
  et~al.}{2021a}]{LIGO_2021_GWTC3}
Abbott R.,  et~al., 2021a

\bibitem[\protect\citeauthoryear{Abbott et~al.,}{Abbott
  et~al.}{2021b}]{Abbott_2021_GWTC2}
Abbott R.,  et~al., 2021b, \mn@doi [\apj] {10.3847/2041-8213/abe949}, 913, L7

\bibitem[\protect\citeauthoryear{Ajith et~al.,}{Ajith
  et~al.}{2007}]{Ajith_2007}
Ajith P.,  et~al., 2007, \mn@doi [CQGra] {10.1088/0264-9381/24/19/s31}, 24,
  S689

\bibitem[\protect\citeauthoryear{Ajith et~al.,}{Ajith
  et~al.}{2011}]{Ajith_etal_2011}
Ajith P.,  et~al., 2011, \mn@doi [PhRvL] {10.1103/PhysRevLett.106.241101}, 106,
  241101

\bibitem[\protect\citeauthoryear{Akiyama et~al.}{Akiyama
  et~al.}{2019}]{EventHorizonTelescope_2019I}
Akiyama K.,  et~al., 2019, \mn@doi [ApJ] {10.3847/2041-8213/ab0ec7}, 875, L1

\bibitem[\protect\citeauthoryear{{Amaro-Seoane} et~al.,}{{Amaro-Seoane}
  et~al.}{2017}]{2017eLISA}
{Amaro-Seoane} P.,  et~al., 2017, arXiv:1702.00786, \href
  {https://ui.adsabs.harvard.edu/abs/2017arXiv170200786A} {}

\bibitem[\protect\citeauthoryear{Antonini, Barausse  \& Silk}{Antonini
  et~al.}{2015}]{Antonini_2015}
Antonini F.,  Barausse E.,   Silk J.,  2015, \mn@doi [ApJ]
  {10.1088/0004-637x/812/1/72}, 812, 72

\bibitem[\protect\citeauthoryear{{Armitage} \& {Natarajan}}{{Armitage} \&
  {Natarajan}}{2002}]{Armitage2002}
{Armitage} P.~J.,  {Natarajan} P.,  2002, \mn@doi [ApJ] {10.1086/339770}, \href
  {https://ui.adsabs.harvard.edu/abs/2002ApJ...567L...9A} {567, L9}

\bibitem[\protect\citeauthoryear{Arzoumanian et~al.}{Arzoumanian
  et~al.}{2020}]{NANOGrav_12d5}
Arzoumanian Z.,  et~al., 2020, \mn@doi [ApJ] {10.3847/2041-8213/abd401}, 905,
  L34

\bibitem[\protect\citeauthoryear{Babak et~al.,}{Babak
  et~al.}{2017}]{Babak:2017tow}
Babak S.,  et~al., 2017, \mn@doi [PhRvD] {10.1103/PhysRevD.95.103012}, 95,
  103012

\bibitem[\protect\citeauthoryear{Baldry et~al.,}{Baldry
  et~al.}{2012}]{Baldry2012}
Baldry I.~K.,  et~al., 2012, \mn@doi [MNRAS]
  {10.1111/j.1365-2966.2012.20340.x}, 421, 621

\bibitem[\protect\citeauthoryear{Barack \& Cutler}{Barack \&
  Cutler}{2004}]{Barack:2003fp}
Barack L.,  Cutler C.,  2004, \mn@doi [PhRvD] {10.1103/PhysRevD.69.082005}, 69,
  082005

\bibitem[\protect\citeauthoryear{Barausse, Dvorkin, Tremmel, Volonteri  \&
  Bonetti}{Barausse et~al.}{2020}]{Barausse_2020}
Barausse E.,  Dvorkin I.,  Tremmel M.,  Volonteri M.,   Bonetti M.,  2020,
  \mn@doi [ApJ] {10.3847/1538-4357/abba7f}, 904, 16

\bibitem[\protect\citeauthoryear{{Begelman}, {Blandford}  \& {Rees}}{{Begelman}
  et~al.}{1980}]{Begelman1980Natur}
{Begelman} M.~C.,  {Blandford} R.~D.,   {Rees} M.~J.,  1980, \mn@doi [\nat]
  {10.1038/287307a0}, \href
  {https://ui.adsabs.harvard.edu/abs/1980Natur.287..307B} {287, 307}

\bibitem[\protect\citeauthoryear{Bhagwat, Pacilio, Barausse  \& Pani}{Bhagwat
  et~al.}{2022}]{Bhagwat:2021kwv}
Bhagwat S.,  Pacilio C.,  Barausse E.,   Pani P.,  2022, \mn@doi [PhRvD]
  {10.1103/PhysRevD.105.124063}, 105, 124063

\bibitem[\protect\citeauthoryear{Bhatta}{Bhatta}{2019}]{Bhatta_2019}
Bhatta G.,  2019, \mn@doi [MNRAS] {10.1093/mnras/stz1482}, 487, 3990

\bibitem[\protect\citeauthoryear{Bonetti, Sesana, Barausse  \& Haardt}{Bonetti
  et~al.}{2018}]{Bonetti:2017lnj}
Bonetti M.,  Sesana A.,  Barausse E.,   Haardt F.,  2018, \mn@doi [MNRAS]
  {10.1093/mnras/sty874}, 477, 2599

\bibitem[\protect\citeauthoryear{Bonetti, Sesana, Haardt, Barausse  \&
  Colpi}{Bonetti et~al.}{2019b}]{Bonetti:2018tpf}
Bonetti M.,  Sesana A.,  Haardt F.,  Barausse E.,   Colpi M.,  2019b, \mn@doi
  [MNRAS] {10.1093/mnras/stz903}, 486, 4044

\bibitem[\protect\citeauthoryear{Bonetti, Sesana, Haardt, Barausse  \&
  Colpi}{Bonetti et~al.}{2019a}]{Bonetti2019}
Bonetti M.,  Sesana A.,  Haardt F.,  Barausse E.,   Colpi M.,  2019a, \mn@doi
  [MNRAS] {10.1093/mnras/stz903}, 486, 4044

\bibitem[\protect\citeauthoryear{Breiding, Burke-Spolaor, Eracleous,
  Bogdanovi{\'{c}}, Lazio, Runnoe  \& Sigurdsson}{Breiding
  et~al.}{2021}]{Breiding_2021}
Breiding P.,  Burke-Spolaor S.,  Eracleous M.,  Bogdanovi{\'{c}} T.,  Lazio T.
  J.~W.,  Runnoe J.,   Sigurdsson S.,  2021, \mn@doi [ApJ]
  {10.3847/1538-4357/abfa9a}, 914, 37

\bibitem[\protect\citeauthoryear{Britzen et~al.,}{Britzen
  et~al.}{2018}]{Britzen_2018}
Britzen S.,  et~al., 2018, \mn@doi [MNRAS] {10.1093/mnras/sty1026}, 478, 3199

\bibitem[\protect\citeauthoryear{Callegari, Mayer, Kazantzidis, Colpi,
  Governato, Quinn  \& Wadsley}{Callegari et~al.}{2009}]{Callegari:2008py}
Callegari S.,  Mayer L.,  Kazantzidis S.,  Colpi M.,  Governato F.,  Quinn T.,
   Wadsley J.,  2009, \mn@doi [ApJ] {10.1088/0004-637X/696/1/L89}, 696, L89

\bibitem[\protect\citeauthoryear{{Callegari}, {Kazantzidis}, {Mayer}, {Colpi},
  {Bellovary}, {Quinn}  \& {Wadsley}}{{Callegari}
  et~al.}{2011}]{Callegari_2011}
{Callegari} S.,  {Kazantzidis} S.,  {Mayer} L.,  {Colpi} M.,  {Bellovary}
  J.~M.,  {Quinn} T.,   {Wadsley} J.,  2011, \mn@doi [ApJ]
  {10.1088/0004-637X/729/2/85}, \href
  {https://ui.adsabs.harvard.edu/abs/2011ApJ...729...85C} {729, 85}

\bibitem[\protect\citeauthoryear{Casteels et~al.,}{Casteels
  et~al.}{2014}]{Casteels_etal_2014_GAMA}
Casteels K. R.~V.,  et~al., 2014, \mn@doi [MNRAS] {10.1093/mnras/stu1799}, 445,
  1157

\bibitem[\protect\citeauthoryear{Chen, Yu  \& Lu}{Chen
  et~al.}{2020}]{Chen_2020}
Chen Y.,  Yu Q.,   Lu Y.,  2020, \mn@doi [ApJ] {10.3847/1538-4357/ab9594}, 897,
  86

\bibitem[\protect\citeauthoryear{Chen, Ricarte  \& Pacucci}{Chen
  et~al.}{2022}]{Chen:2022sae}
Chen H.-Y.,  Ricarte A.,   Pacucci F.,  2022

\bibitem[\protect\citeauthoryear{{Colpi}}{{Colpi}}{2014}]{Colpi2014}
{Colpi} M.,  2014, \mn@doi [\ssr] {10.1007/s11214-014-0067-1}, \href
  {https://ui.adsabs.harvard.edu/abs/2014SSRv..183..189C} {183, 189}

\bibitem[\protect\citeauthoryear{Conselice}{Conselice}{2014}]{Conselice:2014joa}
Conselice C.~J.,  2014, \mn@doi [ARA\&A] {10.1146/annurev-astro-081913-040037},
  52, 291

\bibitem[\protect\citeauthoryear{D'Orazio \& Loeb}{D'Orazio \&
  Loeb}{2018}]{D_Orazio_2018}
D'Orazio D.~J.,  Loeb A.,  2018, \mn@doi [ApJ] {10.3847/1538-4357/aad413}, 863,
  185

\bibitem[\protect\citeauthoryear{{Darg} et~al.,}{{Darg}
  et~al.}{2010}]{SDSS_merger_rate_2010}
{Darg} D.~W.,  et~al., 2010, \mn@doi [\mnras]
  {10.1111/j.1365-2966.2009.15686.x}, \href
  {https://ui.adsabs.harvard.edu/abs/2010MNRAS.401.1043D} {401, 1043}

\bibitem[\protect\citeauthoryear{Davies et~al.,}{Davies
  et~al.}{2018}]{Davies_etal_DEVILS2018}
Davies L. J.~M.,  et~al., 2018, \mn@doi [MNRAS] {10.1093/mnras/sty1553}, 480,
  768

\bibitem[\protect\citeauthoryear{{Dennett-Thorpe}, {Scheuer}, {Laing},
  {Bridle}, {Pooley}  \& {Reich}}{{Dennett-Thorpe} et~al.}{2002}]{Dennett2002}
{Dennett-Thorpe} J.,  {Scheuer} P.~A.~G.,  {Laing} R.~A.,  {Bridle} A.~H.,
  {Pooley} G.~G.,   {Reich} W.,  2002, \mn@doi [\mnras]
  {10.1046/j.1365-8711.2002.05106.x}, \href
  {https://ui.adsabs.harvard.edu/abs/2002MNRAS.330..609D} {330, 609}

\bibitem[\protect\citeauthoryear{Dey et~al.,}{Dey et~al.}{2018}]{Dey_2018}
Dey L.,  et~al., 2018, \mn@doi [ApJ] {10.3847/1538-4357/aadd95}, 866, 11

\bibitem[\protect\citeauthoryear{Di~Matteo, Springel  \& Hernquist}{Di~Matteo
  et~al.}{2005}]{DiMatteo:2005ttp}
Di~Matteo T.,  Springel V.,   Hernquist L.,  2005, \mn@doi [Nature]
  {10.1038/nature03335}, 433, 604

\bibitem[\protect\citeauthoryear{{Ding} et~al.,}{{Ding}
  et~al.}{2020}]{Ding_etal_2020}
{Ding} X.,  et~al., 2020, \mn@doi [ApJ] {10.3847/1538-4357/ab5b90}, \href
  {https://ui.adsabs.harvard.edu/abs/2020ApJ...888...37D} {888, 37}

\bibitem[\protect\citeauthoryear{Dotti, Colpi  \& Haardt}{Dotti
  et~al.}{2006}]{Dotti_2006_nucleardisc}
Dotti M.,  Colpi M.,   Haardt F.,  2006, \mn@doi [MNRAS]
  {10.1111/j.1365-2966.2005.09956.x}, 367, 103

\bibitem[\protect\citeauthoryear{{Dotti}, {Sesana}  \& {Decarli}}{{Dotti}
  et~al.}{2012}]{Dotti2012}
{Dotti} M.,  {Sesana} A.,   {Decarli} R.,  2012, \mn@doi [Advances in
  Astronomy] {10.1155/2012/940568}, \href
  {https://ui.adsabs.harvard.edu/abs/2012AdAst2012E...3D} {2012, 940568}

\bibitem[\protect\citeauthoryear{{Dressler} \& {Richstone}}{{Dressler} \&
  {Richstone}}{1988}]{Dressler1988ApJ}
{Dressler} A.,  {Richstone} D.~O.,  1988, \mn@doi [ApJ] {10.1086/165930}, \href
  {https://ui.adsabs.harvard.edu/abs/1988ApJ...324..701D} {324, 701}

\bibitem[\protect\citeauthoryear{Driver et~al.,}{Driver
  et~al.}{2022}]{Driver_etal_2022GAMA}
Driver S.~P.,  et~al., 2022, \mn@doi [MNRAS] {10.1093/mnras/stac472}, 513, 439

\bibitem[\protect\citeauthoryear{Dullo \& Graham}{Dullo \&
  Graham}{2014}]{Dullo_2014}
Dullo B.~T.,  Graham A.~W.,  2014, \mn@doi [MNRAS] {10.1093/mnras/stu1590},
  444, 2700

\bibitem[\protect\citeauthoryear{Escala, Larson, Coppi  \& Mardones}{Escala
  et~al.}{2005}]{Escala_2005}
Escala A.,  Larson R.~B.,  Coppi P.~S.,   Mardones D.,  2005, \mn@doi [ApJ]
  {10.1086/431747}, 630, 152

\bibitem[\protect\citeauthoryear{{Faber} \& {Jackson}}{{Faber} \&
  {Jackson}}{1976}]{Faber_Jackson_1976}
{Faber} S.~M.,  {Jackson} R.~E.,  1976, \mn@doi [ApJ] {10.1086/154215}, \href
  {https://ui.adsabs.harvard.edu/abs/1976ApJ...204..668F} {204, 668}

\bibitem[\protect\citeauthoryear{Fang \& Yang}{Fang \&
  Yang}{2022}]{Fang:2021xab}
Fang Y.,  Yang H.,  2022, \mn@doi [ApJ] {10.3847/1538-4357/ac4bd7}, 927, 93

\bibitem[\protect\citeauthoryear{Fang \& Yang}{Fang \&
  Yang}{2023}]{fang_SMBHB_delay}
Fang Y.,  Yang H.,  2023, \url{https://github.com/YunFang1/SMBHB-merger-delay}

\bibitem[\protect\citeauthoryear{Ferrarese \& Merritt}{Ferrarese \&
  Merritt}{2000}]{Ferrarese_2000}
Ferrarese L.,  Merritt D.,  2000, \mn@doi [ApJ] {10.1086/312838}, 539, L9

\bibitem[\protect\citeauthoryear{{Foreman-Mackey}, {Hogg}, {Lang}  \&
  {Goodman}}{{Foreman-Mackey} et~al.}{2013}]{Foreman2013emcee}
{Foreman-Mackey} D.,  {Hogg} D.~W.,  {Lang} D.,   {Goodman} J.,  2013, \mn@doi
  [\pasp] {10.1086/670067}, \href
  {https://ui.adsabs.harvard.edu/abs/2013PASP..125..306F} {125, 306}

\bibitem[\protect\citeauthoryear{Fu et~al.,}{Fu
  et~al.}{2011a}]{fu2011kiloparsec}
Fu H.,  et~al., 2011a, ApJ, 740, L44

\bibitem[\protect\citeauthoryear{Fu, Yan, Myers, Stockton, Djorgovski, Aldering
   \& Rich}{Fu et~al.}{2011b}]{Fu_2011}
Fu H.,  Yan L.,  Myers A.~D.,  Stockton A.,  Djorgovski S.~G.,  Aldering G.,
  Rich J.~A.,  2011b, \mn@doi [ApJ] {10.1088/0004-637x/745/1/67}, 745, 67

\bibitem[\protect\citeauthoryear{Gebhardt et~al.,}{Gebhardt
  et~al.}{2000}]{Gebhardt_2000}
Gebhardt K.,  et~al., 2000, ] {10.1086/312840}, 539, L13

\bibitem[\protect\citeauthoryear{Goicovic, Sesana, Cuadra  \&
  Stasyszyn}{Goicovic et~al.}{2017}]{Goicovic:2016dul}
Goicovic F.~G.,  Sesana A.,  Cuadra J.,   Stasyszyn F.,  2017, \mn@doi [MNRAS]
  {10.1093/mnras/stx1996}, 472, 514

\bibitem[\protect\citeauthoryear{Graham et~al.,}{Graham
  et~al.}{2015a}]{Grahammnras1726}
Graham M.~J.,  et~al., 2015a, \mn@doi [MNRAS] {10.1093/mnras/stv1726}, 453,
  1562

\bibitem[\protect\citeauthoryear{Graham et~al.,}{Graham
  et~al.}{2015b}]{graham2015possible}
Graham M.~J.,  et~al., 2015b, Nature, 518, 74

\bibitem[\protect\citeauthoryear{Granato, Zotti, Silva, Bressan  \&
  Danese}{Granato et~al.}{2004}]{Granato_2004}
Granato G.~L.,  Zotti G.~D.,  Silva L.,  Bressan A.,   Danese L.,  2004,
  \mn@doi [ApJ] {10.1086/379875}, 600, 580

\bibitem[\protect\citeauthoryear{{Grazian} et~al.,}{{Grazian}
  et~al.}{2015}]{Grazian2015AA}
{Grazian} A.,  et~al., 2015, \mn@doi [\aap] {10.1051/0004-6361/201424750},
  \href {https://ui.adsabs.harvard.edu/abs/2015A&A...575A..96G} {575, A96}

\bibitem[\protect\citeauthoryear{Gualandris, Read, Dehnen  \&
  Bortolas}{Gualandris et~al.}{2016}]{Gualandris2016}
Gualandris A.,  Read J.~I.,  Dehnen W.,   Bortolas E.,  2016, \mn@doi [MNRAS]
  {10.1093/mnras/stw2528}, 464, 2301

\bibitem[\protect\citeauthoryear{{Haehnelt} \& {Kauffmann}}{{Haehnelt} \&
  {Kauffmann}}{2002}]{Haehnelt2002}
{Haehnelt} M.~G.,  {Kauffmann} G.,  2002, \mn@doi [\mnras]
  {10.1046/j.1365-8711.2002.06056.x}, \href
  {https://ui.adsabs.harvard.edu/abs/2002MNRAS.336L..61H} {336, L61}

\bibitem[\protect\citeauthoryear{Haiman, Kocsis  \& Menou}{Haiman
  et~al.}{2009}]{Haiman_2009}
Haiman Z.,  Kocsis B.,   Menou K.,  2009, \mn@doi [ApJ]
  {10.1088/0004-637x/700/2/1952}, 700, 1952

\bibitem[\protect\citeauthoryear{{Hobbs} et~al.,}{{Hobbs}
  et~al.}{2010}]{Hobbs_etal_2010}
{Hobbs} G.,  et~al., 2010, \mn@doi [CQGra] {10.1088/0264-9381/27/8/084013},
  \href {https://ui.adsabs.harvard.edu/abs/2010CQGra..27h4013H} {27, 084013}

\bibitem[\protect\citeauthoryear{Hoffman \& Loeb}{Hoffman \&
  Loeb}{2007a}]{Hoffman_Loeb2007}
Hoffman L.,  Loeb A.,  2007a, \mn@doi [MNRAS]
  {10.1111/j.1365-2966.2007.11694.x}, 377, 957

\bibitem[\protect\citeauthoryear{Hoffman \& Loeb}{Hoffman \&
  Loeb}{2007b}]{Hoffman_Loeb_2007}
Hoffman L.,  Loeb A.,  2007b, \mn@doi [MNRAS]
  {10.1111/j.1365-2966.2007.11694.x}, 377, 957

\bibitem[\protect\citeauthoryear{Holley-Bockelmann \&
  Sigurdsson}{Holley-Bockelmann \&
  Sigurdsson}{2006}]{Holley-Bockelmann:2006gbs}
Holley-Bockelmann K.,  Sigurdsson S.,  2006

\bibitem[\protect\citeauthoryear{Hopkins, Hernquist, Cox, Matteo, Robertson  \&
  Springel}{Hopkins et~al.}{2006}]{Hopkins_2006}
Hopkins P.~F.,  Hernquist L.,  Cox T.~J.,  Matteo T.~D.,  Robertson B.,
  Springel V.,  2006, \mn@doi [ApJ Supplement Series] {10.1086/499298}, 163, 1

\bibitem[\protect\citeauthoryear{Hu \& Wu}{Hu \& Wu}{2017}]{YueliangWu_taiji}
Hu W.-R.,  Wu Y.-L.,  2017, \mn@doi [National Science Review]
  {10.1093/nsr/nwx116}, 4, 685

\bibitem[\protect\citeauthoryear{{Hu{\v{s}}ko}, {Lacey}  \&
  {Baugh}}{{Hu{\v{s}}ko} et~al.}{2022}]{Filip_2022_galaxy}
{Hu{\v{s}}ko} F.,  {Lacey} C.~G.,   {Baugh} C.~M.,  2022, \mn@doi [\mnras]
  {10.1093/mnras/stab3324}, \href
  {https://ui.adsabs.harvard.edu/abs/2022MNRAS.509.5918H} {509, 5918}

\bibitem[\protect\citeauthoryear{Jenet, Lommen, Larson  \& Wen}{Jenet
  et~al.}{2004}]{Jenet_2004}
Jenet F.~A.,  Lommen A.,  Larson S.~L.,   Wen L.,  2004, \mn@doi [ApJ]
  {10.1086/383020}, 606, 799

\bibitem[\protect\citeauthoryear{{Jiang} et~al.,}{{Jiang}
  et~al.}{2022}]{Jiang_Yang_2022}
{Jiang} N.,  et~al., 2022, arXiv:2201.11633

\bibitem[\protect\citeauthoryear{Kass \& Raftery}{Kass \&
  Raftery}{1995}]{Kass_Raftery_1995}
Kass R.~E.,  Raftery A.~E.,  1995, \mn@doi [Journal of the American Statistical
  Association] {10.1080/01621459.1995.10476572}, 90, 773

\bibitem[\protect\citeauthoryear{{Kelley}, {Charisi}, {Burke-Spolaor}  \& {et
  al}}{{Kelley} et~al.}{2019}]{NANOGrav:2019tvo}
{Kelley} L.,  {Charisi} M.,  {Burke-Spolaor} S.,   {et al} 2019, Bulletin of
  the American Astronomical Society, \href
  {https://ui.adsabs.harvard.edu/abs/2019BAAS...51c.490K} {51, 490}

\bibitem[\protect\citeauthoryear{{Khan}, {Just}  \& {Merritt}}{{Khan}
  et~al.}{2011}]{Khan_2011}
{Khan} F.~M.,  {Just} A.,   {Merritt} D.,  2011, \mn@doi [ApJ]
  {10.1088/0004-637X/732/2/89}, \href
  {https://ui.adsabs.harvard.edu/abs/2011ApJ...732...89K} {732, 89}

\bibitem[\protect\citeauthoryear{{Kharb}, {Lal}  \& {Merritt}}{{Kharb}
  et~al.}{2017}]{Kharb_2017}
{Kharb} P.,  {Lal} D.~V.,   {Merritt} D.,  2017, \mn@doi [Nature Astronomy]
  {10.1038/s41550-017-0256-4}, \href
  {https://ui.adsabs.harvard.edu/abs/2017NatAs...1..727K} {1, 727}

\bibitem[\protect\citeauthoryear{Klein et~al.}{Klein
  et~al.}{2016}]{Klein:2015hvg}
Klein A.,  et~al., 2016, \mn@doi [PhRvD] {10.1103/PhysRevD.93.024003}, 93,
  024003

\bibitem[\protect\citeauthoryear{{Komossa}, {Burwitz}, {Hasinger}, {Predehl},
  {Kaastra}  \& {Ikebe}}{{Komossa} et~al.}{2003}]{Komossa2003_NGC6240}
{Komossa} S.,  {Burwitz} V.,  {Hasinger} G.,  {Predehl} P.,  {Kaastra} J.~S.,
  {Ikebe} Y.,  2003, \mn@doi [ApJ] {10.1086/346145}, \href
  {https://ui.adsabs.harvard.edu/abs/2003ApJ...582L..15K} {582, L15}

\bibitem[\protect\citeauthoryear{{Kormendy}}{{Kormendy}}{1993}]{Kormendy1993nag}
{Kormendy} J.,  1993, in {Beckman} J.,  {Colina} L.,   {Netzer} H.,  eds, The
  Nearest Active Galaxies. pp 197--218

\bibitem[\protect\citeauthoryear{Kormendy \& Ho}{Kormendy \&
  Ho}{2013}]{Kormendy:2013dxa}
Kormendy J.,  Ho L.~C.,  2013, \mn@doi [ARA&A]
  {10.1146/annurev-astro-082708-101811}, 51, 511

\bibitem[\protect\citeauthoryear{{Lauer} et~al.,}{{Lauer}
  et~al.}{2007}]{Lauer_etal_2007}
{Lauer} T.~R.,  et~al., 2007, \mn@doi [ApJ] {10.1086/518223}, \href
  {https://ui.adsabs.harvard.edu/abs/2007ApJ...662..808L} {662, 808}

\bibitem[\protect\citeauthoryear{Liu, Li  \& Komossa}{Liu
  et~al.}{2014}]{Liu_2014}
Liu F.~K.,  Li S.,   Komossa S.,  2014, \mn@doi [ApJ]
  {10.1088/0004-637x/786/2/103}, 786, 103

\bibitem[\protect\citeauthoryear{Lotz, Jonsson, Cox, Croton, Primack,
  Somerville  \& Stewart}{Lotz et~al.}{2011}]{Lotz_2011}
Lotz J.~M.,  Jonsson P.,  Cox T.~J.,  Croton D.,  Primack J.~R.,  Somerville
  R.~S.,   Stewart K.,  2011, \mn@doi [ApJ] {10.1088/0004-637x/742/2/103}, 742,
  103

\bibitem[\protect\citeauthoryear{Luo et~al.,}{Luo
  et~al.}{2016}]{luo2016tianqin}
Luo J.,  et~al., 2016, CQGra, 33, 035010

\bibitem[\protect\citeauthoryear{Magorrian et~al.,}{Magorrian
  et~al.}{1998}]{Magorrian_1998}
Magorrian J.,  et~al., 1998, \mn@doi [AJ] {10.1086/300353}, 115, 2285

\bibitem[\protect\citeauthoryear{Mayer, Kazantzidis, Madau, Colpi, Quinn  \&
  Wadsley}{Mayer et~al.}{2007}]{Mayer:2007vk}
Mayer L.,  Kazantzidis S.,  Madau P.,  Colpi M.,  Quinn T.~R.,   Wadsley J.,
  2007, \mn@doi [Science] {10.1126/science.1141858}, 316, 1874

\bibitem[\protect\citeauthoryear{Mazzolari, Bonetti, Sesana, Colombo, Dotti,
  Lodato  \& Izquierdo-Villalba}{Mazzolari et~al.}{2022}]{Mazzolari:2022cho}
Mazzolari G.,  Bonetti M.,  Sesana A.,  Colombo R.~M.,  Dotti M.,  Lodato G.,
  Izquierdo-Villalba D.,  2022, ] {10.1093/mnras/stac2255}

\bibitem[\protect\citeauthoryear{{McLeod}, {McLure}, {Dunlop}, {Cullen},
  {Carnall}  \& {Duncan}}{{McLeod} et~al.}{2021}]{McLeod2021}
{McLeod} D.~J.,  {McLure} R.~J.,  {Dunlop} J.~S.,  {Cullen} F.,  {Carnall}
  A.~C.,   {Duncan} K.,  2021, \mn@doi [\mnras] {10.1093/mnras/stab731}, \href
  {https://ui.adsabs.harvard.edu/abs/2021MNRAS.503.4413M} {503, 4413}

\bibitem[\protect\citeauthoryear{Merritt \& Ekers}{Merritt \&
  Ekers}{2002}]{Merritt:2002hc}
Merritt D.,  Ekers R.~D.,  2002, \mn@doi [Science] {10.1126/science.1074688},
  297, 1310

\bibitem[\protect\citeauthoryear{{Merritt} \& {Milosavljevi{\'c}}}{{Merritt} \&
  {Milosavljevi{\'c}}}{2005}]{Merritt2005review}
{Merritt} D.,  {Milosavljevi{\'c}} M.,  2005, \mn@doi [Living Reviews in
  Relativity] {10.12942/lrr-2005-8}, \href
  {https://ui.adsabs.harvard.edu/abs/2005LRR.....8....8M} {8, 8}

\bibitem[\protect\citeauthoryear{Merritt \& Poon}{Merritt \&
  Poon}{2004}]{Merritt_2004}
Merritt D.,  Poon M.~Y.,  2004, \mn@doi [ApJ] {10.1086/382497}, 606, 788

\bibitem[\protect\citeauthoryear{Merritt \& Szell}{Merritt \&
  Szell}{2006}]{Merritt_2006}
Merritt D.,  Szell A.,  2006, \mn@doi [ApJ] {10.1086/506010}, 648, 890

\bibitem[\protect\citeauthoryear{Middleton, Sesana, Chen, Vecchio, Del~Pozzo
  \& Rosado}{Middleton et~al.}{2021}]{Middleton:2020asl}
Middleton H.,  Sesana A.,  Chen S.,  Vecchio A.,  Del~Pozzo W.,   Rosado P.~A.,
   2021, \mn@doi [MNRAS] {10.1093/mnrasl/slab008}, 502, L99

\bibitem[\protect\citeauthoryear{Milosavljevi{\'{c}} \&
  Merritt}{Milosavljevi{\'{c}} \& Merritt}{2001}]{Milosavljevi__2001}
Milosavljevi{\'{c}} M.,  Merritt D.,  2001, \mn@doi [ApJ] {10.1086/323830},
  563, 34

\bibitem[\protect\citeauthoryear{{Milosavljevi{\'c}} \&
  {Merritt}}{{Milosavljevi{\'c}} \& {Merritt}}{2003a}]{Milosavljevic2003}
{Milosavljevi{\'c}} M.,  {Merritt} D.,  2003a, \mn@doi [ApJ] {10.1086/378086},
  \href {https://ui.adsabs.harvard.edu/abs/2003ApJ...596..860M} {596, 860}

\bibitem[\protect\citeauthoryear{{Milosavljevi{\'c}} \&
  {Merritt}}{{Milosavljevi{\'c}} \& {Merritt}}{2003b}]{Milosavljevic2003b}
{Milosavljevi{\'c}} M.,  {Merritt} D.,  2003b, in {Centrella} J.~M.,  ed.,
  American Institute of Physics Conference Series Vol. 686, The Astrophysics of
  Gravitational Wave Sources. pp 201--210 (\mn@eprint {arXiv}
  {astro-ph/0212270}), \mn@doi{10.1063/1.1629432}

\bibitem[\protect\citeauthoryear{Milosavljević, Merritt, Rest  \& van~den
  Bosch}{Milosavljević et~al.}{2002}]{Milosavljevic_2002}
Milosavljević M.,  Merritt D.,  Rest A.,   van~den Bosch F.~C.,  2002, \mn@doi
  [MNRAS] {10.1046/j.1365-8711.2002.05436.x}, 331, L51

\bibitem[\protect\citeauthoryear{{Mundy}, {Conselice}, {Duncan}, {Almaini},
  {H{\"a}u{\ss}ler}  \& {Hartley}}{{Mundy} et~al.}{2017}]{Mundy_2017}
{Mundy} C.~J.,  {Conselice} C.~J.,  {Duncan} K.~J.,  {Almaini} O.,
  {H{\"a}u{\ss}ler} B.,   {Hartley} W.~G.,  2017, \mn@doi [\mnras]
  {10.1093/mnras/stx1238}, \href
  {https://ui.adsabs.harvard.edu/abs/2017MNRAS.470.3507M} {470, 3507}

\bibitem[\protect\citeauthoryear{Naoz, Rose, Michaely, Melchor, Ramirez-Ruiz,
  Mockler  \& Schnittman}{Naoz et~al.}{2022}]{Naoz:2022rru}
Naoz S.,  Rose S.~C.,  Michaely E.,  Melchor D.,  Ramirez-Ruiz E.,  Mockler B.,
    Schnittman J.~D.,  2022, \mn@doi [ApJ] {10.3847/2041-8213/ac574b}, 927, L18

\bibitem[\protect\citeauthoryear{{O'Leary}, {Moster}, {Naab}  \&
  {Somerville}}{{O'Leary} et~al.}{2021}]{OLeary_2021}
{O'Leary} J.~A.,  {Moster} B.~P.,  {Naab} T.,   {Somerville} R.~S.,  2021,
  \mn@doi [\mnras] {10.1093/mnras/staa3746}, \href
  {https://ui.adsabs.harvard.edu/abs/2021MNRAS.501.3215O} {501, 3215}

\bibitem[\protect\citeauthoryear{Pan \& Yang}{Pan \& Yang}{2021}]{Pan:2021ksp}
Pan Z.,  Yang H.,  2021, \mn@doi [PhRvD] {10.1103/PhysRevD.103.103018}, 103,
  103018

\bibitem[\protect\citeauthoryear{Pan, Lyu  \& Yang}{Pan
  et~al.}{2021}]{Pan:2021oob}
Pan Z.,  Lyu Z.,   Yang H.,  2021, \mn@doi [PhRvD]
  {10.1103/PhysRevD.104.063007}, 104, 063007

\bibitem[\protect\citeauthoryear{Perley, Chandler, Butler  \& Wrobel}{Perley
  et~al.}{2011}]{Perley_2011}
Perley R.~A.,  Chandler C.~J.,  Butler B.~J.,   Wrobel J.~M.,  2011, \mn@doi
  [ApJ] {10.1088/2041-8205/739/1/l1}, 739, L1

\bibitem[\protect\citeauthoryear{Reines \& Volonteri}{Reines \&
  Volonteri}{2015}]{Reines_2015}
Reines A.~E.,  Volonteri M.,  2015, \mn@doi [ApJ] {10.1088/0004-637x/813/2/82},
  813, 82

\bibitem[\protect\citeauthoryear{Robson, Cornish  \& Liu}{Robson
  et~al.}{2019}]{Robson:2018ifk}
Robson T.,  Cornish N.~J.,   Liu C.,  2019, \mn@doi [CQGra]
  {10.1088/1361-6382/ab1101}, 36, 105011

\bibitem[\protect\citeauthoryear{Rodriguez-Gomez et~al.}{Rodriguez-Gomez
  et~al.}{2015a}]{Rodriguez-Gomez:2015aua}
Rodriguez-Gomez V.,  et~al., 2015a, \mn@doi [MNRAS] {10.1093/mnras/stv264},
  449, 49

\bibitem[\protect\citeauthoryear{Rodriguez-Gomez et~al.,}{Rodriguez-Gomez
  et~al.}{2015b}]{Vicente2015}
Rodriguez-Gomez V.,  et~al., 2015b, \mn@doi [MNRAS] {10.1093/mnras/stv264},
  449, 49

\bibitem[\protect\citeauthoryear{Rodriguez, Taylor, Zavala, Peck, Pollack  \&
  Romani}{Rodriguez et~al.}{2006}]{Rodriguez:2006th}
Rodriguez C.,  Taylor G.~B.,  Zavala R.~T.,  Peck A.~B.,  Pollack L.~K.,
  Romani R.~W.,  2006, \mn@doi [ApJ] {10.1086/504825}, 646, 49

\bibitem[\protect\citeauthoryear{{Schweizer}}{{Schweizer}}{1996}]{Schweizer1996book}
{Schweizer} F.,  1996, in , Galaxies: Interactions and Induced Star Formation
  Saas-Fee Advanced Course 26 Lecture Notes 1996. Kennicutt.
pp 105--274

\bibitem[\protect\citeauthoryear{Seoane et~al.}{Seoane
  et~al.}{2013}]{eLISA:2013}
Seoane P.~A.,  et~al., 2013

\bibitem[\protect\citeauthoryear{Stefanon, Bouwens, Labb{\'{e}}, Illingworth,
  Gonzalez  \& Oesch}{Stefanon et~al.}{2021}]{Stefanon_2021}
Stefanon M.,  Bouwens R.~J.,  Labb{\'{e}} I.,  Illingworth G.~D.,  Gonzalez V.,
    Oesch P.~A.,  2021, \mn@doi [ApJ] {10.3847/1538-4357/ac1bb6}, 922, 29

\bibitem[\protect\citeauthoryear{{Thatte} et~al.,}{{Thatte}
  et~al.}{2021}]{Thatte_etal_2021}
{Thatte} N.,  et~al., 2021, \mn@doi [The Messenger] {10.18727/0722-6691/5215},
  \href {https://ui.adsabs.harvard.edu/abs/2021Msngr.182....7T} {182, 7}

\bibitem[\protect\citeauthoryear{Toubiana, Wong, Babak, Barausse, Berti, Gair,
  Marsat  \& Taylor}{Toubiana et~al.}{2021}]{Toubiana:2021iuw}
Toubiana A.,  Wong K. W.~K.,  Babak S.,  Barausse E.,  Berti E.,  Gair J.~R.,
  Marsat S.,   Taylor S.~R.,  2021, \mn@doi [PhRvD]
  {10.1103/PhysRevD.104.083027}, 104, 083027

\bibitem[\protect\citeauthoryear{{Vasiliev}}{{Vasiliev}}{2016}]{Vasiliev2016}
{Vasiliev} E.,  2016, in {Meiron} Y.,  {Li} S.,  {Liu} F.~K.,   {Spurzem} R.,
  eds,  Vol. 312, Star Clusters and Black Holes in Galaxies across Cosmic Time.
  pp 92--100 (\mn@eprint {arXiv} {1411.1762}),
  \mn@doi{10.1017/S1743921315007607}

\bibitem[\protect\citeauthoryear{Vasiliev, Antonini  \& Merritt}{Vasiliev
  et~al.}{2015}]{Vasiliev_2015}
Vasiliev E.,  Antonini F.,   Merritt D.,  2015, \mn@doi [ApJ]
  {10.1088/0004-637x/810/1/49}, 810, 49

\bibitem[\protect\citeauthoryear{Volonteri \& Natarajan}{Volonteri \&
  Natarajan}{2009}]{Volonteri2009_Msigma}
Volonteri M.,  Natarajan P.,  2009, \mn@doi [MNRAS]
  {10.1111/j.1365-2966.2009.15577.x}, 400, 1911

\bibitem[\protect\citeauthoryear{Volonteri, Haardt  \& Madau}{Volonteri
  et~al.}{2003}]{Volonteri_2003}
Volonteri M.,  Haardt F.,   Madau P.,  2003, \mn@doi [ApJ] {10.1086/344675},
  582, 559

\bibitem[\protect\citeauthoryear{Wang \& Gao}{Wang \&
  Gao}{2010}]{Wang_2010_Xray}
Wang J.-B.,  Gao Y.,  2010, \mn@doi [Res. Astron. Astrophys.]
  {10.1088/1674-4527/10/4/003}, 10, 309

\bibitem[\protect\citeauthoryear{Weisskopf, Tananbaum, van Speybroeck  \&
  O'Dell}{Weisskopf et~al.}{2000}]{Weisskopf_2000_Chandra}
Weisskopf M.~C.,  Tananbaum H.~D.,  van Speybroeck L.~P.,   O'Dell S.~L.,
  2000, \mn@doi [Proc. SPIE Int. Soc. Opt. Eng.] {10.1117/12.391545}, 4012, 2

\bibitem[\protect\citeauthoryear{Xu, Zhao, Scoville, Capak, Drory  \& Gao}{Xu
  et~al.}{2012}]{Xu_2012}
Xu C.~K.,  Zhao Y.,  Scoville N.,  Capak P.,  Drory N.,   Gao Y.,  2012,
  \mn@doi [ApJ] {10.1088/0004-637x/747/2/85}, 747, 85

\bibitem[\protect\citeauthoryear{{Yu}}{{Yu}}{2002}]{Yu2002}
{Yu} Q.,  2002, \mn@doi [\mnras] {10.1046/j.1365-8711.2002.05242.x}, \href
  {https://ui.adsabs.harvard.edu/abs/2002MNRAS.331..935Y} {331, 935}

\bibitem[\protect\citeauthoryear{Yunes, Coleman~Miller  \& Thornburg}{Yunes
  et~al.}{2011}]{Yunes:2010sm}
Yunes N.,  Coleman~Miller M.,   Thornburg J.,  2011, \mn@doi [PhRvD]
  {10.1103/PhysRevD.83.044030}, 83, 044030

\bibitem[\protect\citeauthoryear{Zhang \& Jin}{Zhang \&
  Jin}{1997}]{Zhang_Jin_1996}
Zhang S.,  Jin J.,  1997, American Journal of Physics, 65, 355

\makeatother
\end{thebibliography}
%
\bsp
\label{lastpage}
\end{document}